\documentclass[twocolumn]{openjournal}
\usepackage[utf8]{inputenc}
\accepted{September 17, 2025}
\shortauthors{Polanska et al.}

\usepackage{graphicx}
\usepackage{amsmath}
\usepackage{xcolor}
\usepackage[caption=false]{subfig}
\usepackage{comment}
\usepackage{verbatim}
\usepackage{hyperref}
\usepackage{fontawesome5}
\usepackage{booktabs}

\usepackage{soul}

\hypersetup{
    colorlinks=true,
    linkcolor=blue,
    filecolor=blue,
    urlcolor=blue,
    citecolor=blue,
    }

\newcommand{\prob}{\ensuremath{{p}}}

\newcommand{\data}{\ensuremath{{y}}}
\newcommand{\datum}{\ensuremath{y}}
\renewcommand{\exp}{\ensuremath{\text{exp}}}
\newcommand{\given}{\ensuremath{{\,\vert\,}}}

\begin{document}

\title{Learned harmonic mean estimation of the Bayesian evidence\\ with normalizing flows
}

\author{Alicja Polanska$^{1\star}$}
\author{Matthew A. Price$^{1}$}
\author{Davide Piras$^{2,3}$}
\author{Alessio Spurio Mancini$^{4,1}$}
\author{Jason D. McEwen$^{1,5\dagger}$}

\thanks{$^\star$ E-mail: \href{mailto:alicja.polanska.22@ucl.ac.uk}{alicja.polanska.22@ucl.ac.uk}}
\thanks{$^\dagger$ E-mail: \href{mailto:jason.mcewen@ucl.ac.uk}{jason.mcewen@ucl.ac.uk}}

\affiliation{$^{1}$ Mullard Space Science Laboratory, University College London, Dorking, RH5 6NT, UK}
\affiliation{$^{2}$ Centre Universitaire d'Informatique, Université de Genève, 1227 Genève 4, Switzerland}
\affiliation{$^{3}$ Département de Physique Théorique, Université de Genève, 24 quai Ernest Ansermet, 1211 Genève 4, Switzerland}
\affiliation {$^{4}$ Department of Physics, Royal Holloway, University of London, Egham Hill, Egham, UK}
\affiliation{$^{5}$ Alan Turing Institute, London, NW1 2DB, UK}

\begin{abstract}
  We present the learned harmonic mean estimator with normalizing flows -- a robust, scalable and flexible estimator of the Bayesian evidence for model comparison.  Since the estimator is agnostic to sampling strategy and simply requires posterior samples, it can be applied to compute the evidence using any Markov chain Monte Carlo (MCMC) sampling technique, including saved down MCMC chains, or any variational inference approach.  The learned harmonic mean estimator was recently introduced, where machine learning techniques were developed to learn a suitable internal importance sampling target distribution to solve the issue of exploding variance of the original harmonic mean estimator.  In this article we present the use of normalizing flows as the internal machine learning technique within the learned harmonic mean estimator.  Normalizing flows can be elegantly coupled with the learned harmonic mean to provide an approach that is more robust, flexible and scalable than the machine learning models considered previously. We perform a series of numerical experiments, applying our method to benchmark problems and to a cosmological example in up to $21$ dimensions. We find the learned harmonic mean estimator is in agreement with ground truth values and nested sampling estimates.  The open-source \texttt{harmonic} Python package implementing the learned harmonic mean, now with normalizing flows included, is publicly available. \href{https://github.com/astro-informatics/harmonic/}{\faGithub}
\end{abstract}

\section{Introduction}
\label{sec:introduction}
Model selection plays a crucial role in understanding the complexities of the Universe. It involves the task of identifying the underlying model that best describes observations, for instance of astrophysical phenomena. The field of Bayesian statistics provides a framework for statistical inference and decision-making that incorporates prior knowledge to update probabilities based on observed data. This approach is well-suited for cosmology, for example, as experiments in the field tend to consist of single observations of events, as opposed to repeatable experiments which are at the core of the frequentist framework. As a consequence, Bayesian inference and model comparison are widespread in the field \citep{trotta2008bayes}. In the Bayesian formalism, an essential tool in this process is the estimation of the Bayesian evidence, also called the marginal likelihood, which quantifies the probability of observed data given a model. The Bayesian evidence allows us to evaluate the relative plausibility of models and assess which hypotheses are best supported by the available data, which is of course not only useful in cosmology but in many other fields.

As a topical illustration of the importance of model selection in cosmology, recent baryon acoustic oscillations  measurements from the Dark Energy Spectroscopic Instrument \citep{desicollaboration2016desi}, combined with observations of the cosmic microwave background \citep{aghanim2020planck, carron2022cmb, madhavacheril2024atacama} and with supernovae Ia measurements from PantheonPlus \citep{Brout_2022}, Union3 \citep{rubin2023union} or DESY5 \citep{descollaboration2024dark}, provide a tantalizing suggestion of the existence of a time-varying dark energy equation-of-state.  Whether dark energy can be described by Einstein's cosmological constant or whether an equation-of-state with $w \neq -1$ is required is a fundamental question of modern cosmology that we hope to answer definitively in the near future through the application of Bayesian model selection techniques to upcoming observational data.  We showcase the application of the methodology presented in this article to precisely this question through a simulated Dark Energy Survey (DES) galaxy clustering and weak lensing analysis \citep[cf.][]{abbott2018dark}.

In practice, the computation of Bayesian evidence is very challenging as it involves evaluating a multi-dimensional integral over a potentially highly varied function. The most widespread method for estimating the Bayesian evidence, particularly in astrophysics, is nested sampling \citep{skilling2006nested,ashton2022nested,Buchner2021}.  While nested sampling has been highly successful and many effective nested sampling algorithms and codes have been developed \citep{Feroz_2008, 2009MNRAS.398.1601F,Feroz_2009, 2010ascl.soft10029B, Handley_2015a,Handley_2015,Feroz_2019,2020MNRAS.493.3132S,Buchner2021, Williams_2021, cai2022proximal}, it imposes constraints on the method used to sample.  By sampling in a nested manner it is possible to reparameterize the likelihood in terms of the enclosed prior volume such that the evidence can be computed by a one-dimensional integral.  The computational challenge then shifts to how to effectively sample in a nested manner, i.e.\ how to sample from the prior subject to likelihood level-sets or isocontours.  The need to sample in this nested manner severely reduces flexibility  (hence the need to design custom nested sampling algorithms), typically restricting application to relatively low-dimensional settings.\footnote{A notable exception that is applicable to high-dimensional settings is proximal nested sampling \citep{cai2022proximal}, although it is only applicable for log-convex likelihoods.}

The harmonic mean estimator of the Bayesian evidence, introduced by \citet{newton1994approximate}, provides much greater flexibility since it only requires samples from the posterior, available from any Markov chain Monte Carlo (MCMC) method, for example. However, it was immediately realized by \citet{neal:1994} that the method can easily fail catastrophically due to the estimator's variance becoming very large.  To solve this issue the learned harmonic mean estimator was recently proposed by \citet{mcewen2023machine}, where machine learning techniques were developed to learn a suitable internal importance sampling target distribution. Other evidence estimation methods decoupled from the evidence have been proposed recently \citep{heavens2017marginal, pmlr-v118-jia20a, srinivasan2024floz}.
Since the estimator requires only samples from the posterior and so is agnostic to the method used to generate samples, in contrast to nested sampling, it can be easily applied with any MCMC sampling technique, including saved down MCMC chains, or any variational inference approach.  This property also allows the estimator to be adapted to address Bayesian model selection for simulation-based inference (SBI) \citep{Spurio_Mancini_2023}, where an explicit likelihood is unavailable or infeasible.

In this article we present the use of normalizing flows as the internal machine learning technique within the learned harmonic mean estimator.  Normalizing flows can be elegantly coupled with the learned harmonic mean to provide an approach that is more robust, flexible and scalable than the machine learning models considered previously.  In \citet{polanska2023learned} we presented preliminary work introducing normalizing flow as the machine learning technique within the learned harmonic mean.  We fully develop the methodology in the current article, introduce the use of additional, more expressive flows, and perform more extensive numerical experiments validating and showcasing the method.  The \texttt{harmonic}\footnote{\url{https://github.com/astro-informatics/harmonic}} Python package implementing the learned harmonic mean estimator, including with normalizing flows, is publicly available.

While normalizing flows can learn a normalized posterior density by definition, the normalization constant itself, \textit{i.e.}\ the Bayesian evidence, is not directly accessible.  Nevertheless, the Bayesian evidence can be computed by backing out the normalization constant, as discussed in \cite{Spurio_Mancini_2023}, by taking the ratio of the unnormalized posterior (given by the product of the likelihood and prior) with the normalizing flow representing a surrogate for the posterior.   This approach, which we call the na\"ive normalizing flow estimator in \cite{Spurio_Mancini_2023}, is highly dependent on the accuracy of the approximating normalizing flow and suffers a large variance, as discussed in \cite{Spurio_Mancini_2023}.  For comparison, we compute this na\"ive estimator in the current article and demonstrate its large variance.
Very recently, \citet{srinivasan2024floz} adopt this na\"ive estimator and attempt to reduce its variance by introducing an additional term in the loss that penalizes variability when training the flow.  While this reduces the variability of the estimator, the estimator nevertheless remains highly dependent on the accuracy of the approximating flow and there is no statistical guarantee that resulting evidence estimates are unbiased.
Training flows using the forward Kullback-Leibler (KL) divergence when given samples from the target distribution is known to suffer from mode covering behaviour, where the learned flow has wider tails than the target \citep[e.g.][]{pml1Book}.  While the approach presented in \citet{srinivasan2024floz} suffers from this problem which can directly impact the accuracy of estimated evidence values, our learned harmonic mean estimator does not.  Firstly, in the learned harmonic mean approach with normalizing flow presented in this article, we concentrate the probability density of the internal importance sampling target distribution that is learned. Secondly, the distribution that is learned in our approach is in any case not used as a surrogate for the posterior so it need not be an accurate approximation. For further details see Section~\ref{sec:learned_harmonic_mean} or \citealp{mcewen2023machine}.

The remainder of this article is structured as follows. In Section~\ref{sec:harmonic_mean} we briefly review Bayesian model comparison, the original harmonic mean estimator, elucidating its catastrophic failure arising from its large variance, and the learned harmonic mean estimator, which solves this large variance problem.  In Section~\ref{sec:learned_harmonic_mean_flows_big} we describe normalizing flows and how they can be integrated elegantly into the learned harmonic mean framework to provide a more robust, flexible and scalable approach than the simple machine learning models considered previously. In Section~\ref{sec:experiments} we present numerical experiments that validate the effectiveness of our method.  This includes low-dimensional benchmark examples where the ground truth value is accessible and a higher-dimensional practical cosmological example on DES-like simulations, as discussed above, where we validate against the evidence value computed by nested sampling. Finally, in Section~\ref{sec:conclusions} we present concluding remarks.

\section{The harmonic mean estimator}
In this section we briefly review Bayesian model comparison, the original harmonic mean estimator, and the learned harmonic mean estimator.  We discuss the exploding variance problem of the original harmonic mean and describe how the learned harmonic mean solves this problem.

\label{sec:harmonic_mean}
\subsection{Bayesian model comparison}
\label{sec:bayesian_model_comparison}
Using empirical data to test theoretical models lies at the heart of the scientific method, the foundation of research progress and innovation. Bayesian model comparison is a powerful approach for evaluating the relative plausibility of competing models in the light of data. In the Bayesian framework probability distributions provide a quantification of uncertainty.

Bayes' theorem is a fundamental principle in Bayesian statistics that allows us to update our beliefs about models in light of observed data. Consider observed data $\data$ described through a model $M$ parametrised by $\theta$. Bayes' theorem gives us the posterior $\prob(\theta \given \data, M)$, the probability density of a model's parameter $\theta$ given observed data $\data$ and model $M$. It is expressed in terms of the prior probability density of the model, the likelihood of the data under that model, and the Bayesian evidence for the data:
\begin{equation}
  \label{eq:bayes}
  \prob(\theta \given \data, M)
  = \frac{\prob(\data \given \theta, M) \prob(\theta \given M)}{\prob(\data \given M)}
  = \frac{\mathcal{L}(\theta) \pi(\theta)}{z}.
\end{equation}
The likelihood $\prob(\data \given \theta, M) = \mathcal{L}(\theta)$ expresses how probable the observed data $y$ is for different values of the parameter $\theta$. The prior $\prob(\theta \given M) = \pi(\theta)$ quantifies pre-existing knowledge or assumptions about $\theta$. The Bayesian evidence, also called the marginal likelihood, $\prob(\data \given M) = z$ is a normalizing factor for the posterior distribution.

The Bayesian evidence is often omitted when estimating parameters, for instance using MCMC methods, as only the relative values of the posterior probability are of interest. However, it is a crucial quantity in Bayesian model comparison. It quantifies the probability of observing the data under a particular model, integrating over the model's parameter space:
\begin{equation}
  \label{eqn:evidence}
  z =
  \prob(\data \given M)
  = \int \,\text{d} \theta \
  \prob(\data \given \theta, M) \prob(\theta \given M)
  = \int \,\text{d} \theta \
  \mathcal{L}(\theta) \pi(\theta).
\end{equation}
The Bayesian evidence can be used to compute Bayes' factors to provide a direct measure of the relative support for one model over another. The Bayes' factor between two models $M_{1}$, $M_{2}$ is defined as
\begin{equation}
  \text{BF}_{12} = \frac{\prob(\data \given M_{1})}{\prob(\data \given M_{2})}.
\end{equation}
Given prior model probabilities, Bayes' factors offer a straightforward way to compare models and help make informed decisions about model selection.

In practice, the Bayesian evidence can be very challenging to calculate as $\theta$ is often high-dimensional. As a result, computing $z$ involves evaluating a multi-dimensional integral over a potentially highly varied function. In principle, this could be done through a standard MCMC integration of the posterior, but this approach is not accurate in practice, even in relative low dimensions. Many alternative methods have been proposed; for reviews see \citet{friel2012estimating,clyde2007current}.
The most popular method for computing the evidence, particularly in the astrophysics community, is nested sampling \citet{skilling2006nested}.  As discussed already, many highly effective nested sampling algorithms have been developed.  However, nested sampling imposes strong constraints on the method used to generate samples, significantly reducing its flexibility.  Consequently, custom nested sampling algorithms must be designed and are typically restricted to relatively low dimensional settings.

\subsection{The original harmonic mean estimator}
\label{sec:original_harmonic_mean}
The original harmonic mean estimator of the Bayesian evidence was introduced by \citet{newton1994approximate}, providing an expression for the reciprocal Bayesian evidence $\rho = z^{-1}$ given by
\begin{align}
  \rho & = \label{eq:rho_exp}
  \mathbb{E}_{\prob(\theta \given \data)} \biggl[
    \frac{1}{\mathcal{L}(\theta)}
    \biggr].
\end{align}
This motivates the harmonic mean estimator $\hat{\rho}$ of the reciprocal Bayesian evidence, which can be written as an expectation of the reciprocal of the likelihood under the posterior,
\begin{equation}
  \hat{\rho} = \frac{1}{N} \sum_{i=1}^{N} \frac{1}{\mathcal{L}(\theta_i)} ,
  \quad
  \theta_i \sim \prob(\theta \given \data).
\end{equation}
The Bayesian evidence can then be straightforwardly obtained as the inverse $\hat{z} = \hat{\rho}^{-1}$ (although a more accurate estimator of the evidence from its reciprocal can also be considered; \citealt{mcewen2023machine}). In principle, this estimator provides a simple and flexible method of evaluating the Bayesian evidence.

However, it was quickly realised that the harmonic mean estimator can be highly inaccurate due to its variance growing very large \citep{neal:1994, clyde2007current,friel2012estimating}. The reason for this can be seen when interpreting the harmonic mean estimator through the lens of importance sampling \citep[e.g.][]{mcewen2023machine}. Equation \eqref{eq:rho_exp} can be rewritten as
\begin{align}
  \rho= \int \,\text{d} \theta \:
  \frac{1}{z} \:
  \frac{\pi(\theta)}{\prob(\theta \given \data)} \:
  \prob(\theta \given \data).
\end{align}
It is clear that this expectation is equivalent to importance sampling, where the target density is the prior $\pi(\theta)$ and the sampling density is the posterior $\prob(\theta \given \data)$. This is in contrast to the typical importance sampling use case, where the posterior is the target distribution. For importance sampling to be effective, the sampling density must have fatter tails than the target in order for the target parameter space to be explored efficiently. If this condition is not fulfilled, the variance of the expectation becomes large. In the case of the harmonic mean estimator, the target density (prior) will normally have fatter tails than the sampling density (posterior). This is because the posterior gets updated with new information about the model encoded in the data, and as a result becomes narrower. Thus, the original harmonic mean estimator suffers from an exploding variance issue and is often inaccurate.

\subsection{Learned harmonic mean estimator}
\label{sec:learned_harmonic_mean}
One strategy to remedy the exploding variance problem of the harmonic mean estimator was proposed by \citet{gelfand1994bayesian}, where an arbitrary normalized density $\varphi(\theta)$ is introduced to rewrite the expectation in Equation \eqref{eq:rho_exp} as
\begin{equation}
  \rho
  =
  \mathbb{E}_{\prob(\theta | \data)} \biggl[
    \frac{\varphi(\theta)}{\mathcal{L}(\theta) \pi(\theta)}
    \biggr],
\end{equation}
which naturally results in the estimator
\begin{equation}
  \label{eqn:harmonic_mean_retargeted}
  \hat{\rho} =
  \frac{1}{N} \sum_{i=1}^N
  \frac{\varphi(\theta_i)}{\mathcal{L}(\theta_i) \pi(\theta_i)} ,
  \quad
  \theta_i \sim \prob(\theta | \data).
\end{equation}
The density $\varphi(\theta)$ now takes the role of the importance sampling target. This estimator can therefore remedy the exploding variance problem provided that the target density $\varphi(\theta)$ is selected so that it is contained within the posterior. However, this condition is not trivial to enforce, especially in high dimensions since there is a trade-off between accuracy and efficiency. The contribution to the estimator from each posterior sample $\theta_i$ is weighted by the target density $\varphi(\theta_i)$.  Low weights reduce the contribution of the posterior sample to the estimator, reducing its effective sample size and thus efficiency.  However, the alternative of avoiding low weights can result in a target $\varphi(\theta)$ that is not contained within the posterior, giving rise to the exploding variance problem.  In prior work, a multivariate Gaussian has been considered \citet{gelfand1994bayesian}, although this often fails to contain $\varphi(\theta)$ within the posterior \citep{chib:1995,clyde2007current}.  Indicator functions have also been considered \citep{robert:2009,vanhaasteren:2014}, although typically result in low efficiency. Other solutions to this problem have been proposed but they can be inaccurate, inefficient or limited in their use cases \citep{chib:1995, lenk:2009,raftery:2006}.

The learned harmonic mean estimator was proposed recently by some of the authors of the current article \citep{mcewen2023machine}, where machine learning methods are used to solve the exploding variance problem of the original harmonic mean.
It was realized by \citet{mcewen2023machine} that the optimal target density is the normalized posterior, i.e.
\begin{equation}\label{eq:target}
  \varphi^{\mathrm{optimal}} ({\theta})
  = \frac{\mathcal{L}(\theta) \pi(\theta)}{z}.
\end{equation}
By definition of the problem, however, the normalized posterior is not accessible as the normalization factor is the Bayesian evidence itself. However, the target does not need to be a close approximation of the posterior for the estimate to be correct. It is more important for the target's probability mass to be contained within the posterior to avoid the variance becoming large.
\citet{mcewen2023machine} develop a bespoke optimization approach that learns the posterior density from its samples while ensuring that the resulting model satisfies this condition. They also derive an unbiased estimator of the variance of the estimator and the variance of the variance, which are empirically shown to be accurate. The estimate of the Bayesian evidence computed with the learned harmonic mean thus comes with an error estimate, which can give an indication of how confident one should be in the result, and a number of additional sanity checks \citep{mcewen2023machine}.

The learned harmonic mean results in an accurate estimator of the Bayesian evidence that is agnostic to the sampling strategy, just like the original harmonic mean. This property ensures flexibility of the method, meaning it can be used in conjunction with efficient MCMC sampling techniques and variational inference approaches.
The learned harmonic mean has been shown to be highly accurate on numerous example problems, including several cases where the original harmonic mean had been shown to fail catastrophically \citep{clyde2007current,friel2012estimating}.  However, the bespoke training approach requires an appropriate model to be chosen carefully and the hyperparameters to be fine-tuned through cross validation. Moreover, the simple machine learning models considered previously do not scale well to high-dimensional settings.


\section{Learned harmonic mean estimator with normalizing flows}
In this section we describe the learned harmonic mean with normalizing flow for estimation of the Bayesian evidence.  Normalizing flows can be elegantly coupled with the learned harmonic mean to provide an approach that is more robust, flexible and scalable than the machine learning models considered previously.

\label{sec:learned_harmonic_mean_flows_big}
\subsection{Normalizing flows}
\label{sec:flows}

Normalizing flows meet the core requirements of the learned target distribution of the learned harmonic mean estimator: namely, they provide a normalized probability distribution for which one can evaluate probability densities.
Flows are a class of machine learning model, where an underlying probability distribution is learned, e.g., from training data. The learned distribution can then be sampled from, generating new data instances similar to those in the training set. The learned approximation of the probability density is also accessible, and it is normalized, which is crucial for our use.

Normalizing flow models work by transforming a simple base distribution into a more complex distribution through a series of bijections (invertible transformations). For a comprehensive review of normalizing flows we refer the reader to \citet{papamakarios2021normalizing}. The base distribution is chosen so that it is easy to sample from and to evaluate its probability density, typically a Gaussian with unit variance. A vector $\theta$ of an unknown distribution $p(\theta)$, can be expressed through a transformation $B$ of a latent vector $u$ sampled from the base distribution $q(u)$:
\begin{equation}
  \theta = B(u), \text{ where } u \sim q(u).
\end{equation}
$B$ must be invertible and $B$ and its inverse $B^{-1}$ must be differentiable. When these conditions are satisfied, we can simply calculate the density of the distribution of $\theta$ through the change of variables formula by
\begin{equation}
  p(\theta) = q(u) \vert \det J_{B}(u)  \vert^{-1},
\end{equation}
where $J_{B}(u)$ is the Jacobian corresponding to $B$. Such transformations are composable: $p(\theta)$ can be transformed again, and the resulting normalized density can be obtained analogously. In practice $B$ consists of a series of transformations. These are often defined in such a way that the determinant of $J_{B}(u)$ can be computed efficiently. This is where the power of normalizing flows lies -- a simple base distribution, when taken through a series of simple transformations can become much more expressive and is able to approximate complex targets. In reality, the resulting distribution is an imperfect approximation of $p(\theta)$ that we call $p_{\text{NF}}(\theta, \beta)$, where $\beta$ denotes the trainable parameters of the transformations.

A multitude of flow architectures with different strengths have been proposed. In this work (and in the \texttt{harmonic} code), we use real-valued non-volume preserving \citep{dinh2016density} and rational quadratic spline flows \citep{durkan2019neural}. However, any flow model can be integrated into the method, offering greater computational scalability.


\subsubsection{Real non-volume preserving flows}
Real-valued non-volume preserving (real NVP) flows were introduced by \citet{dinh2016density}. Their architecture is relatively simple, consisting of a series of affine coupling layers. Consider the $D$ dimensional input $x$, split into elements up to and following $d$, respectively, $x_{1:d}$ and $x_{d+1:D}$, for $d<D$.  Given input $x$, the output $y$ of an affine couple layer is calculated by
\begin{align}
  y_{1:d} =   & x_{1:d} ;                                                 \\
  y_{d+1:D} = & x_{d+1:D} \odot \exp\bigl(s(x_{1:d})\bigr) +  t(x_{1:d}),
\end{align}
where $\odot$ denotes Hadamard (elementwise) multiplication, and the scale $s$ and translation $t$ are neural networks with trainable parameters. The Jacobian of such a transformation is a lower-triangular matrix, making its determinant efficient to calculate.

\subsubsection{Rational quadratic spline flows}

A more complex and expressive class of flows are rational quadratic spline flows introduced by \citet{durkan2019neural}. The architecture is similar to real NVP flows, but the layers include monotonic splines. These are piecewise functions consisting of multiple segments of monotonic rational quadratics with learned parameters. Given input $x$, the output $y$ of a rational quadratic coupling layer has the form:
\begin{align}
  \alpha_{1:d} =   & \text{Trainable parameters} ; \\
  \alpha_{d+1:D} = & n (x_{1:d});                  \\
  y_{i} =          & g_{\alpha_{i}}(x_{i}),
\end{align}
where $n$ is a neural network and $g_{\alpha_{i}}$ is a spline parametrised by $\alpha_{i}$, with each bin defined by a monotonically-increasing rational-quadratic function.
Such layers are combined with alternating affine transformations to create the normalizing flow. Thanks to their more expressive and sophisticated architecture, rational quadratic spline flows are well-suited to higher dimensional and more complex problems than real NVP flows \citep{durkan2019neural}.

\begin{figure}
  \centering
  \includegraphics[width=\linewidth]{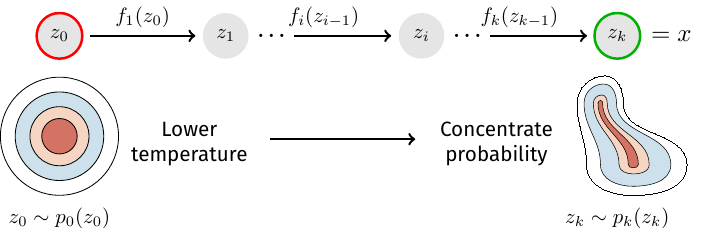}
  \caption{Diagram illustrating how reducing the temperature parameter concentrates the probability density of a normalizing flow. The trained flow at $T=1$ is a normalized approximation of the posterior distribution. The variance of the base distribution, which we call the temperature parameter $T \in (0,1)$, is reduced, concentrating the probability density of the transformed distribution. This ensures that it is contained within the posterior, which is a necessary condition for the internal learned importance target distribution of the learned harmonic mean estimator.}
  \label{fig:temperature_diagram}
\end{figure}


\subsection{The learned harmonic mean estimator with normalizing flows}
\label{sec:learned_harmonic_mean_flows}

In this work we address the limitations of the simple machine learning methods considered in the learned harmonic mean framework previously.  Recall, the aim is to learn an approximation of the posterior from samples but with the critical constraint that the tails of the learned distribution are contained within the posterior.  To learn appropriate models a bespoke optimization algorithm was considered in \citet{mcewen2023machine}.  Normalizing flows afford an elegant alternative solution for keeping the learned target density contained within the posterior, rendering the bespoke training approach unnecessary.  The importance sampling target density is first learned using a normalizing flow model and then concentrated by reducing the variance of the base distribution, i.e.\ reducing its ``temperature''.  The resulting method provides an improved estimator of the Bayesian evidence that retains the flexibility and accuracy of its predecessor, while improving its robustness and scalability.

\subsubsection{Training the flow}

Before we estimate the evidence we need to train a normalizing flow on samples from the posterior. When training normalizing flows, the forward KL divergence is well-suited as the loss function $L$ when we have samples of the target distribution.  Consider an unknown posterior distribution of interest $p(\theta)$ and its approximating flow $p_{\text{NF}}(\theta, \beta)$, where $\beta$ are the trainable flow parameters. The KL divergence can be interpreted as a measure of the dissimilarity between $p(\theta)$ and  $p_{\text{NF}}(\theta, \beta)$ and is therefore a natural quantity to minimize when training a normalizing flow.  The forward KL divergence between the two distributions can be expressed as
\begin{align}
  \label{eq:kl}
  L(\beta) = & D_{KL} \left[\,  p(\theta) \, ||   \, p_{\text{NF}}(\theta, \beta ) \, \right] \nonumber                        \\
  =          & -\mathbb{E}_{p(\theta)} \biggl[ \log \left(\,  p_{\text{NF}}(\theta, \beta ) \, \right) \biggr] + \text{const.}
\end{align}
Given $N$ samples $\theta_{i}$ from the posterior, where $i=\{1, \ldots, N\}$, the expectation in Equation \eqref{eq:kl} can be approximated by Monte Carlo as
\begin{equation}
  \label{eq:loss}
  L(\beta) \approx - \frac{1}{N} \sum_{i=1}^{N} \log \left(\,  p_{\text{NF}}(\theta_{i}, \beta ) \, \right) + \text{const.}
\end{equation}
Minimizing this approximation is equivalent to fitting the normalizing flow to the samples by maximum likelihood \citep{papamakarios2021normalizing}. We take this approach to training our flow, and minimize the loss given by Equation \eqref{eq:loss} using the Adam optimizer \citep{kingma2017adam,dozat.2016}. We use a portion of the samples for training the flow and reserve the rest to be used for inference, to be substituted when estimating the evidence.

It is worth stressing that this is the standard training approach for normalizing flows. By replacing the simple machine learning methods considered in \citet{mcewen2023machine} with flows, we render their bespoke training approach unnecessary, making the method more robust and flexible.

\subsubsection{Concentrating the probability density}
Once the flow is trained on samples from the posterior, we concentrate its probability density by reducing what we call the flow temperature parameter $T$. This is a factor $T \in (0,1)$ by which the variance of the base Gaussian distribution is multiplied. Reducing the base distribution's variance has the effect of concentrating its probability density in parameter space, or reducing its ``temperature'' in a statistical mechanics interpretation. This has the effect of also concentrating the probability density of the transformed distribution due to the continuity and differentiability of the flow, as illustrated in Figure~\ref{fig:temperature_diagram}. Hence, the concentrated flow is the perfect candidate for the importance sampling target in the harmonic mean estimator, as it is normalized and close to the posterior but contained within it. After a flow is trained, it can be used in the learned harmonic mean estimator with different temperature values without the need to retrain for each $T$.

\subsubsection{Standardization}

We standardize the training and inference data.  We calculate the mean and variance of the input training data represented in a matrix $\Theta^{\text{train}}$ and remove that before fitting the model. This means that each entry of the data matrix is transformed as
\begin{equation}
  \Theta^{\text{train}}_{ij} \mapsto (\Theta^{\text{train}}_{ij}-\overline{\Theta}^{\text{train}}_{j})/\sigma^{\text{train}}_{j},
\end{equation}
where $\overline{\Theta}^{\text{train}}_{j}$ is the mean and $\sigma^{\text{train}}_{j}$ is the standard deviation of the training data parameter column $j$ (calculated over the data points). This training data consists of samples from the parameter space so $j \in {1, \ldots, D}$, where $D$ is the dimension of $\theta$. We then apply this same transformation, with $\overline{\Theta}^\text{train}$ and $\sigma^{\text{train}}$ vectors kept the same, to the data points for which we are predicting the probability density, namely the inference data. For the density to still be normalized, we need to then also multiply the flow density by the Jacobian of this transformation, so the predicted density for a standardized model $p^{S}_{\text{NF}}(\theta)$ is
\begin{equation}
  p^{S}_{\text{NF}}(\theta) = p_{\text{NF}}(\theta) \prod_{j=1}^{D} (\sigma^{\text{train}}_{j})^{-1}.
\end{equation}


\subsubsection{Evidence error estimate}
In addition to an estimate of the evidence itself, we also require an estimate of its error.  In \citet{mcewen2023machine} approaches are proposed to estimate the variance of the learned harmonic mean estimator and also its variance.  Specifically, the Bayesian evidence estimate and its error are considered $\hat{\rho} \pm \hat{\sigma}$.  While quoting these terms is sufficient for many toy problems, to ensure numerical stability for practical problems in higher dimensions it is necessary to always work in log space to avoid numerical overflow. Converting the error estimate $\hat{\sigma}$ to log space is non-trivial as $\log(\text{var}(x)) \neq\text{var}(\log(x))$ in general.  To remain in log space we are interested in the log-space error $\hat{\zeta}_{\pm}$ defined by
\begin{equation}
  \log ( \hat{\rho} \pm \hat{\sigma} ) = \log (\hat{\rho}) + \hat{\zeta}_\pm .
\end{equation}
The log-space error estimate can be computed by
\begin{equation}
  \hat{\zeta}_\pm = \log (\hat{\rho} \pm \hat{\sigma} ) - \log (\hat{\rho}) = \log(1 \pm \hat{\sigma} / \hat{\rho}),
\end{equation}
where
\begin{equation}
  \hat{\sigma} / \hat{\rho} =  \exp \bigl( \log(\hat{\sigma}) - \log(\hat{\rho}) \bigr).
\end{equation}
This way we can avoid computing $\hat{\rho} \pm \hat{\sigma}$ directly. We only compute $\log(\hat{\sigma}) - \log(\hat{\rho})$, which we expect to be much smaller and less susceptible to overflow. When quoting the result with log-space errors we use the notation $\log( \hat{\rho})^{ \hat{\zeta}_{+}}_{\hat{\zeta}_{-}}$.  The log evidence errors can be straightforwardly obtained by swapping the negative and positive errors of the reciprocal log evidence.

\subsubsection{Code}
The learned harmonic mean estimator with normalizing flows is implemented in the \texttt{harmonic} package\footnote{\url{https://github.com/astro-informatics/harmonic}}, from version 1.2.0 onwards. The methodology described in this section, with real NVP and rational quadratic spline flows, has been implemented in JAX and is available in recent releases of \texttt{harmonic} on PyPi and GitHub. Furthermore, other parts of the \texttt{harmonic} code have been updated use JAX, which is a Python framework offering acceleration, just-in-time compilation and automatic differentiation functionality \citep{jax2018github}.
Consequently, \texttt{harmonic} can now be run on hardware accelerators such as GPUs, potentially reducing computation times and allowing the user to tackle more complex, computationally demanding problems. Additionally, the automatic differentiation functionality opens up the possibility of optimizing based on evidence (e.g.\ for experimental design), as gradients are now accessible all the way down to evidence level, which provides an intriguing avenue for further research. The normalizing flow portion of the code is implemented using the \texttt{flax} \citep{flax2020github}, TensorFlow Probability \citep{dillon2017tensorflow}, \texttt{optax} and \texttt{distrax} \citep{deepmind2020jax} packages.

\subsubsection{Na\"ive Bayesian evidence estimation using normalizing flows}
\label{sec:naive_estimate}

As discussed in Section~\ref{sec:introduction}, and first described in \citet{Spurio_Mancini_2023}, since flows are normalized it is possible to back out their normalizing constant to provide an estimate of the Bayesian evidence.  We recall this approach here for reference.

Given samples $\theta_i$ an estimate of the evidence can be computed for each sample by
\begin{equation}
  z_i = \frac{\mathcal{L}(\theta_i) \pi(\theta_i)}{p_{\text{NF}}(\theta_i)}.
\end{equation}
While posterior samples are typically available and hence used, in principle the samples $\theta_i$ do not necessarily need to be drawn from the posterior.  An overall estimate of the evidence and its spread can then simply be computed from the mean of these evidence estimates and their standard deviation.  However, the resulting evidence estimator is likely to be biased and will have a large variance.

\section{Numerical experiments}
\label{sec:experiments}
To validate the effectiveness of the method presented in this paper, we perform a series of numerical experiments. Firstly, in Section \ref{sec:benchmarks} we repeat a series of low-dimensional benchmark problems performed by \citet{mcewen2023machine} but using normalizing flows to learn the importance sampling target. The underlying examples are described in more detail by \citet{mcewen2023machine}. The original harmonic mean estimator has been shown to fail catastrophically for many of these examples \citep{friel2012estimating}, while our learned harmonic mean remains accurate.
In Section~\ref{sec:temperature} we study the impact of varying the temperature parameter on the evidence estimate, showing the robustness of our method.
Then in Section \ref{sec:cosmo} we present a practical application of our method in a cosmological context for the DES (Dark Energy Survey). We perform a joint lensing-clustering analysis (``3x2pt'') on a DES Y1-like configuration. We compare our results with the values obtained through the conventional method of nested sampling.


\subsection{Architectures, sampling and training}
In the experiments where a real NVP flow is used, translation networks of the affine coupling layers are given by two-layer dense neural networks with a leaky ReLU activation function in between. For the scaling layers this is scaled by another dense layer with a softplus activation. We permute the inputs between coupling layers to ensure the flow transforms all elements.
In the low dimensional benchmark experiments we consider a real NVP flow, unless otherwise stated, with six coupling layers, where typically only the first two include scaling.
When we use a rational quadratic spline flow, it has a range $-10$ to $10$ (outside of this range it defaults to a linear transformation). The conditioner for the spline hyperparameters is a multi-layer perceptron with a hyperbolic tangent activation. We use a Gaussian base distribution with zero mean and an identity covariance matrix for all flows.

\begin{figure}[t]
  \centering
  \includegraphics[width=\linewidth]{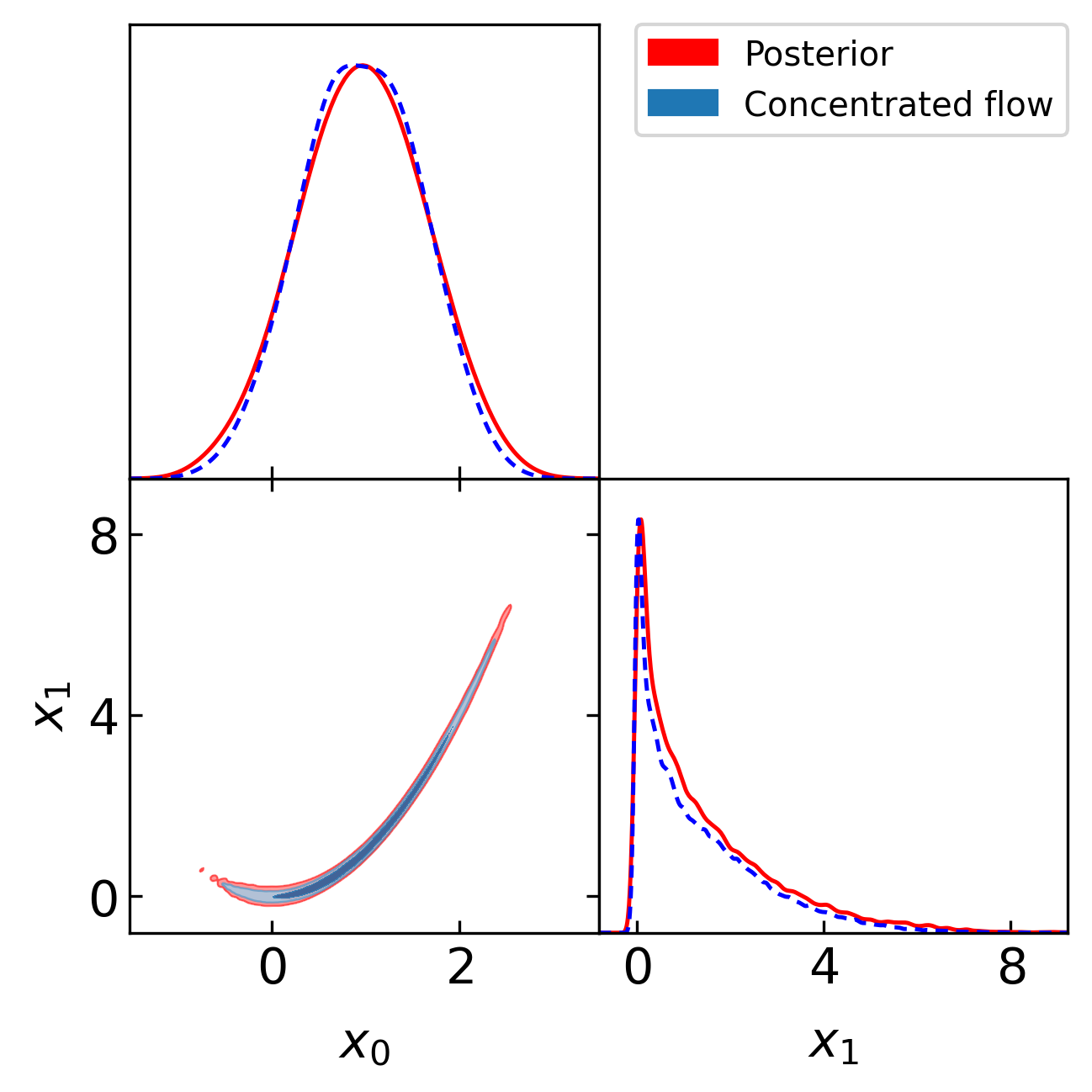}
  \caption{Corner plot of the sampled posterior (solid red) and a real NVP flow with temperature $T=0.9$ (dashed blue) for the Rosenbrock benchmark problem. The internal importance target distribution of the estimator given by the concentrated flow is contained within the posterior, as required for the learned harmonic mean estimator.}
  \label{fig:rosenbrock_corner}
\end{figure}

For the low-dimensional benchmark examples, we generate samples from the posterior using MCMC methods implemented in the \texttt{emcee} package \citep{emcee}. In the practical cosmological example, we use the Metropolis-Hastings sampling approach \citep{metropolis:1953,hastings:1970} implemented in the \texttt{cobaya} package \citep{torrado2019}. We then train the flow on half of the samples by maximum likelihood and use the remaining samples for inference.

\subsection{Benchmark examples}
\label{sec:benchmarks}
\subsubsection{Rosenbrock}
\label{sec:rosenbrock}

\begin{figure}
  \centering
  \subfloat[Reciprocal evidence]{\includegraphics[width=\linewidth]{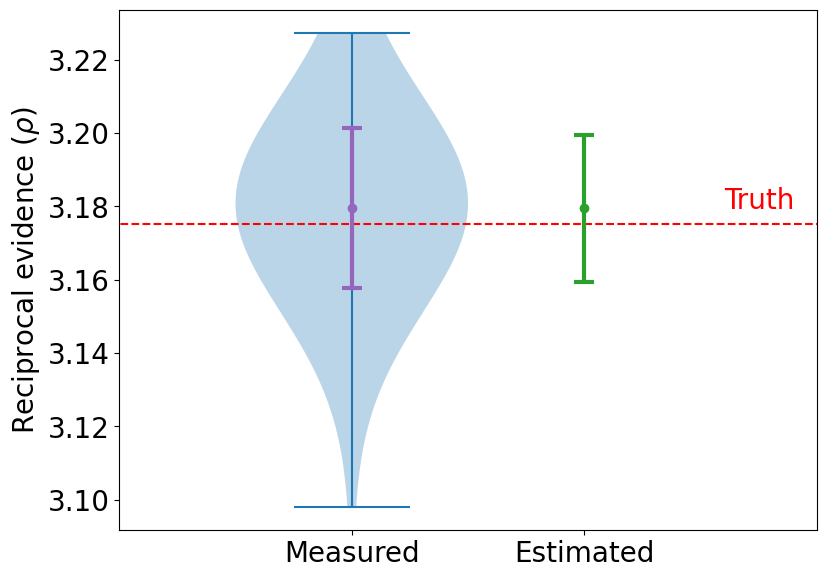} \label{fig:rosenbrock_violin_a}}
  \qquad
  \subfloat[Variance of reciprocal evidence]{\includegraphics[width=\linewidth]{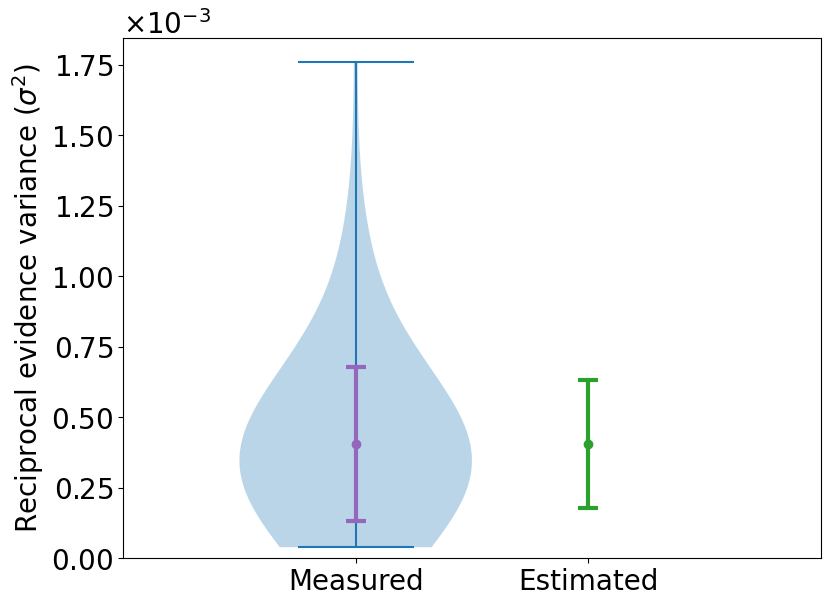} \label{fig:rosenbrock_violin_b}}
  \caption{Violin plots of the reciprocal Bayesian evidence computed by the learned harmonic mean estimator for the Rosenbrock benchmark problem repeated $100$ times. (a) Reciprocal Bayesian evidence estimates across runs (measured) along with the estimate of the standard deviation computed by the error estimator (estimated). The ground truth is shown in red. (b) Sample variance of the estimator across runs (measured) alongside the standard deviation computed by the variance-of-variance estimator (estimated). The evidence estimates and their error estimators are highly accurate.}
  \label{fig:rosenbrock_violin}
\end{figure}

The Rosenbrock problem is a common benchmark example considered when estimating the Bayesian evidence. The Rosenbrock distribution's narrow curving degeneracy presents a challenge in sufficiently exploring the resulting posterior distribution to accurately evaluate the Bayesian evidence. The Rosenbrock function is given by
\begin{equation}
  f({x}) = \sum_{i=1}^{d-1} \bigg [ 100(x_{i+1} - x_{i}^2)^2 + (x_i - 1)^2 \bigg ],
\end{equation}
where $d$ denotes the number of dimensions. In our example we consider a 2-dimensional problem with the log-likelihood given by $\log \mathcal{L}(x) = -f({x})$ and a uniform prior $x_0 \in [-10, 10]$ and $x_1 \in [-5, 15]$.

We draw $1,500$ samples for $200$ chains, with burn-in of $500$ samples, yielding $1,000$ posterior samples per chain. Figure~\ref{fig:rosenbrock_corner} shows the corner plot of samples from the posterior (solid red line) and a real NVP flow with $2$ scaled and $4$ unscaled layers at temperature $T=0.9$ (dashed blue line). It can be seen that the flow approximates the posterior quite well while remaining contained within it. This is exactly what we want in a target distribution for the harmonic mean estimator.

Figure~\ref{fig:rosenbrock_violin_a} shows a violin plot of the results of this experiment repeated $100$ times with posterior samples generated from different seeds. The ground truth obtained through numerical integration is shown in red. It can be seen that the evidence values estimated using our method are accurate, agreeing with the ground truth value. It can also be seen that the estimator of the population variance agrees with the variance measured across the repeats. Figure~\ref{fig:rosenbrock_violin_b} shows a violin plot of the variance estimator across runs alongside the standard deviation calculated from the variance-of-variance estimator. It can be seen they are also in agreement. The Bayesian evidence estimates obtained using the learned harmonic mean and their error estimates are highly accurate.

\subsubsection{Normal-Gamma}

\begin{figure}
  \centering
  \subfloat[Corner plot for the Normal-Gamma example with $\tau_{0}=0.001$]{\includegraphics[width=\linewidth]{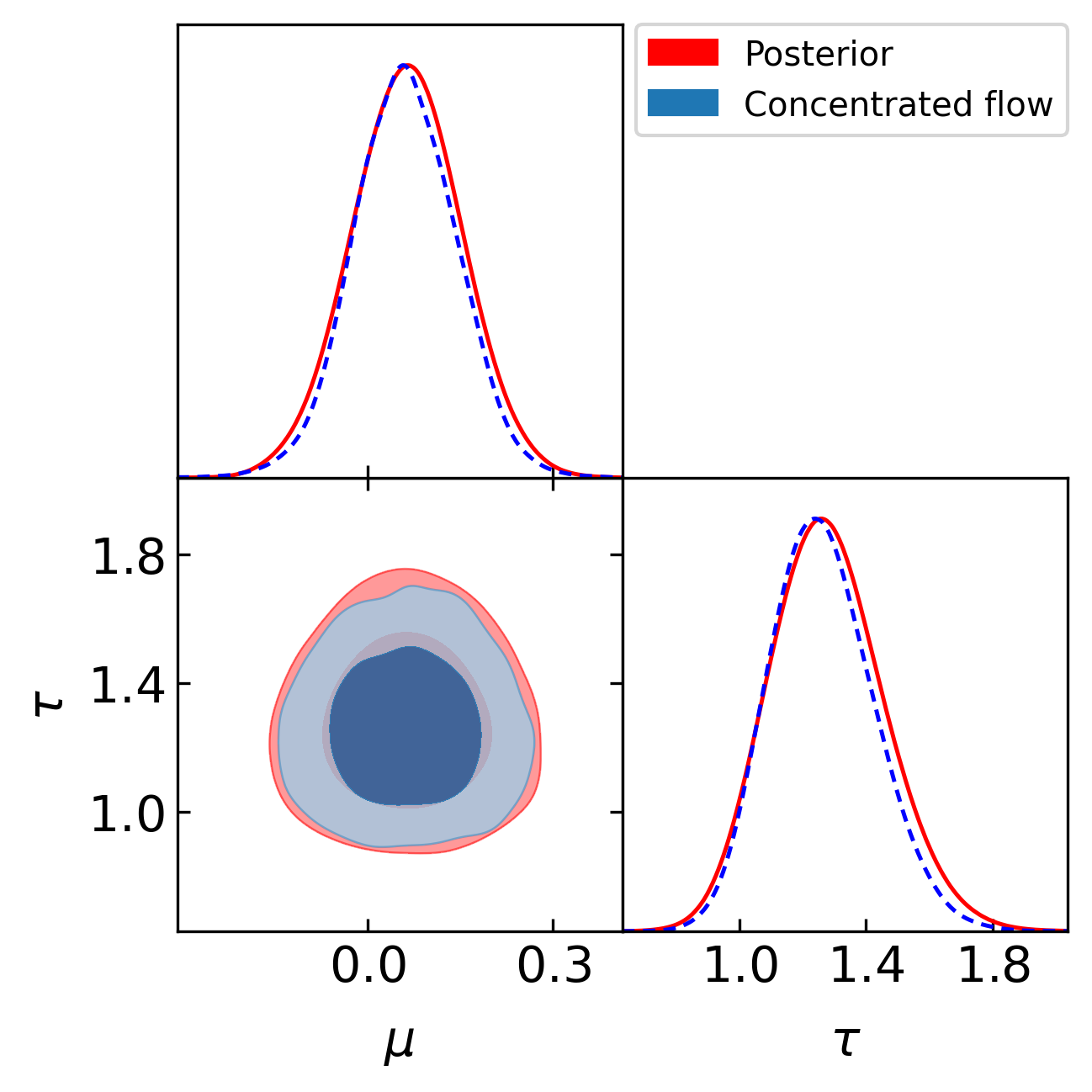} \label{fig:normalgamma_corner}}
  \qquad
  \subfloat[Estimate accuracy for varying prior sizes]{\includegraphics[width=\linewidth]{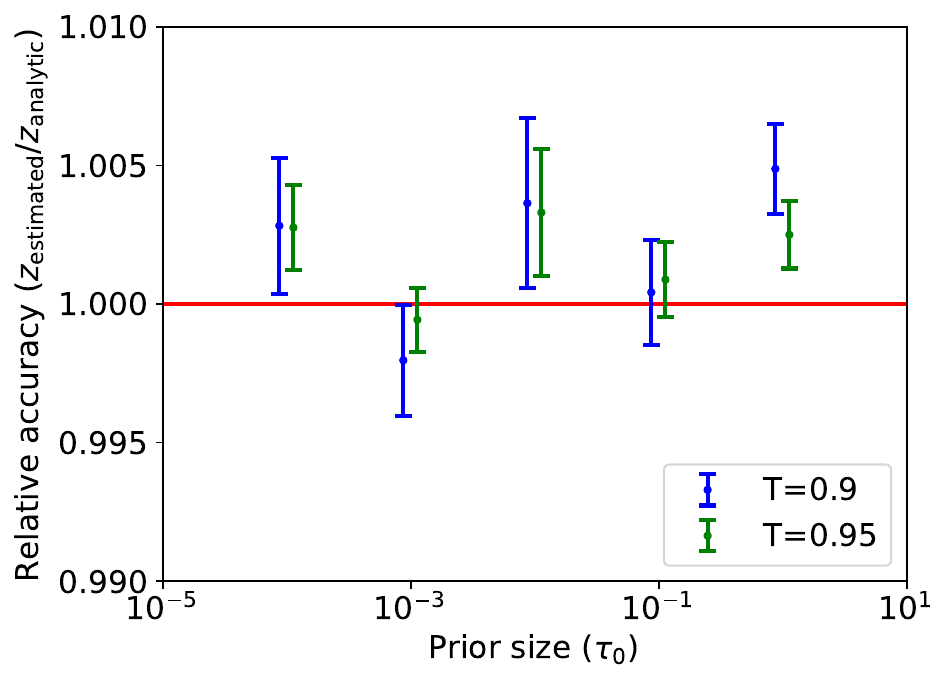} \label{fig:normalgamma_tau}}
  \caption{(a) Corner plot of the sampled posterior (solid red) and real NVP flow with temperature $T=0.9$ (dashed blue) for the Normal-Gamma example with $\tau_{0}=0.001$. The internal importance target distribution given by the concentrated flow is contained within the posterior, as required for the learned harmonic mean estimator. (b) Ratio of Bayesian evidence values computed by the learned harmonic mean estimator with a concentrated flow to those computed analytically for the Normal-Gamma problem with error bars corresponding to the estimated standard deviation. Bayesian evidence estimated with a flow at temperature $T=0.9$ (blue) and $T=0.95$ (green) are shown, with slight offsets for ease of visualization. Unlike the original harmonic mean, our method produces accurate estimates which are sensitive to prior size.}
\end{figure}

We also consider the Normal-Gamma model \citep{bernardo:1994} where data are distributed normally
\begin{equation}
  y_i \sim \text{N}(\mu, \tau^{-1}),
\end{equation}
for $i \in \{1, \ldots, n\}$, with mean $\mu$ and precision $\tau$.  A normal prior is assumed for $\mu$ and a Gamma prior for $\tau$:
\begin{align}
  \mu  & \sim \text{N}\bigl(\mu_0, (\tau_0 \tau)^{-1}\bigr), \\
  \tau & \sim \text{Ga}(a_0, b_0),
\end{align}
with mean $\mu_0 = 0$, shape $a_0 = 10^{-3}$ and rate $b_0 = 10^{-3}$.  The precision scale factor $\tau_0$ controls how diffuse the prior is. \citet{friel2012estimating} apply the original harmonic mean for this example and show that the evidence estimate does not vary with $\tau_{0}$, unlike the analytic ground truth value. We repeat this experiment, drawing $1,500$ samples for $200$ chains, with burn-in of $500$ samples, yielding $1,000$ posterior samples per chain. We use a real NVP flow with $2$ scaled and $4$ unscaled layers at temperatures $T=0.9$ and $T=0.95$ to estimate the evidence.

Figure~\ref{fig:normalgamma_corner} shows an example corner plot of the training samples from the posterior for $\tau=0.001$ (red) and from the normalizing flow (blue) at temperature $T=0.9$. Again, it can be seen that the concentrated learned target is close to the posterior but with thinner tails, as is required.

Figure~\ref{fig:normalgamma_tau} shows the relative accuracy of the evidence estimate computed using our method for a range of prior sizes. It can be seen that, unlike the original harmonic mean estimator, our method is accurate for a range of $\tau_{0}$. Results are computed with the trained flow at temperature $T=0.9$ (blue) and $T=0.95$ (green). They are accurate in both cases, showing that the temperature parameter does not require fine-tuning. A detailed discussion of this point is included in Section~\ref{sec:temperature}.

\vspace{0.5cm}

\subsubsection{Logistic regression models: Pima Indian example}

\begin{figure}
  \centering
  \subfloat[Model 1]{\includegraphics[width=\linewidth]{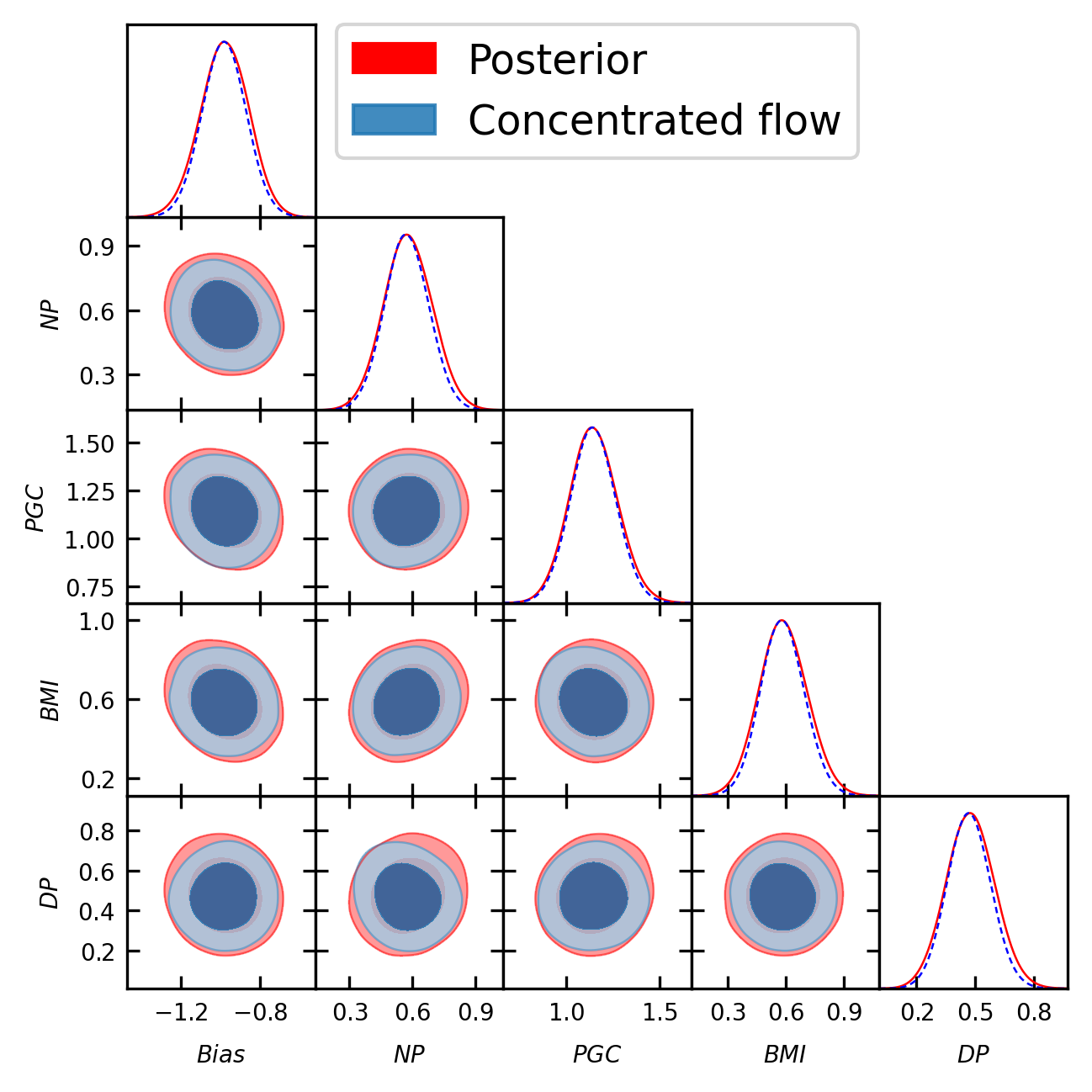} \label{fig:pima_indian_corner_a}}
  \qquad
  \subfloat[Model 2]{\includegraphics[width=\linewidth]{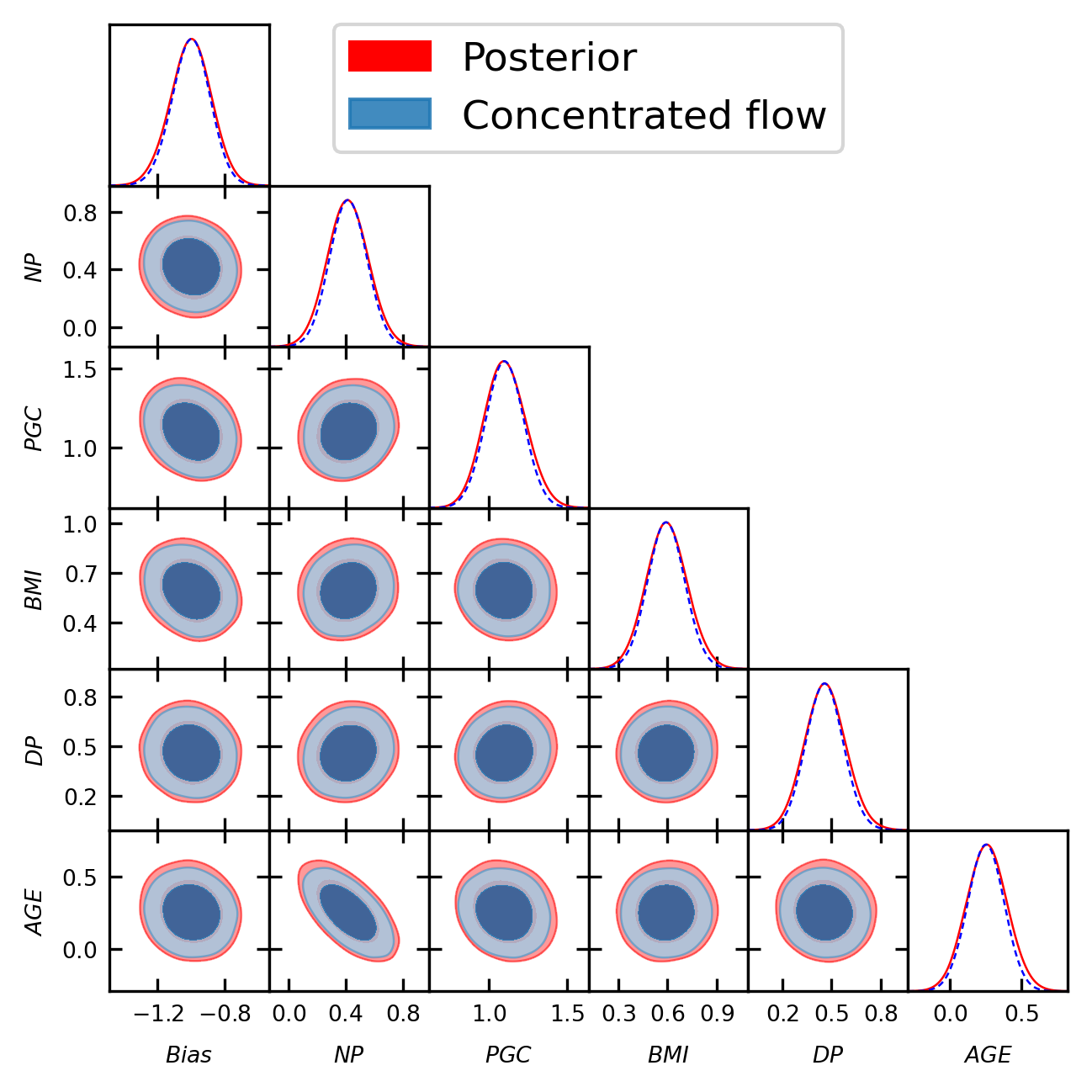} \label{fig:pima_indian_corner_b}}
  \caption{Corner plots of the sampled posterior (solid red) and real NVP flow trained on the posterior samples with temperature $T=0.9$ (dashed blue) for the Pima Indian benchmark problem for $\tau=0.01$. The dimensions correspond to parameters $\theta_{i}$ associated with the covariates included in the analysis.
    The internal importance target distribution given by the concentrated flow is contained within the posterior and has thinner tails, as required for the learned harmonic mean estimator.}
  \label{fig:pima_indian_corner}
\end{figure}

We consider an example involving the comparison of two logistic regression models used to describe the Pima Indians data \citep{smith1988using}, originally from the National Institute of Diabetes and Digestive and Kidney Diseases.

This analysis was originally performed to study indicators of diabetes in $n=532$ Pima Native American women. The predictors of diabetes considered included the number of prior pregnancies (NP), plasma glucose concentration (PGC), body mass index (BMI), diabetes pedigree function (DP) and age (AGE).
The probability of diabetes $p_i$ for person $i \in \{1, \ldots, n\}$ is modelled by the logistic function
\begin{equation}
  p_i = \frac{1}{1+\exp\bigl(- \theta^\text{T} x_i\bigr)},
\end{equation}
with covariates $x_i = (1,x_{i,1}, \dots x_{i,d})^\text{T}$ and parameters $\theta = (\theta_0, \dots, \theta_d)^\text{T}$, where $d$ is the total number of covariates considered.
The likelihood is given by
\begin{equation}
  \mathcal{L}({y} \given {\theta}) = \prod_{i=1}^n p_i^{y_i}(1-p_i)^{1-y_i},
\end{equation}
where $y = (y_1, \dots, y_n)^\text{T}$ is the diabetes incidence with $y_i$ equal to one if patient $i$ has diabetes and zero otherwise. The prior distribution on $\theta$ is a Gaussian with precision $\tau=0.01$.

Two such logistic regression models are considered:
\begin{align*}
  \text{Model } M_1: & \quad \text{covariates = \{NP, PGC, BMI, DP\};}      \\*
  \text{Model } M_2: & \quad \text{covariates = \{NP, PGC, BMI, DP, AGE\},}
\end{align*}
where both additionally include a bias. We estimate the Bayes factor of these models $\text{BF}_{12}$ with our method and compare it to a benchmark value computed by \citet{friel2012estimating} using a reversible jump algorithm \citep{green:1995}. They obtain a value of $\text{BF}_{12}=13.96$ ($\log\text{BF}_{12}=2.636$), which we treat as ground truth. We draw $5,000$ samples for $200$ chains, with burn-in of $1,000$ samples, yielding $4,000$ posterior samples per chain. We use a real NVP flow with $2$ scaled and $4$ unscaled layers at temperature $T = 0.9$, applying standardization.

Figure~\ref{fig:pima_indian_corner} shows the corner plots for this example for both models. The training samples from the posterior are shown in red and from the normalizing flow at temperature $T=0.9$ in blue. Once again, we see that the concentrated flow is contained within the posterior as expected. The log evidence found for Model 1 and 2 is $-257.230^{0.003}_{-0.003}$ and $-259.857^{0.002}_{-0.002}$ respectively, resulting in the estimate $\log\text{BF}_{12}= 2.627^{0.004}_{-0.004}$, indicating a slight preference for Model 1. The Bayes factor value is in close agreement with the benchmark, whereas the original harmonic mean estimator was not accurate \citep{friel2012estimating}.

\subsubsection{Non-nested linear regression models: Radiata pine
  example}
In the last benchmark example we compare two non-nested linear regression models describing the Radiata pine data \citep{williams:1959}. The dataset consists of measurements of the maximum compression strength parallel to the grain $y_i$, density $x_i$ and resin-adjusted density $z_i$, for specimen $i \in \{1, \ldots, n\}$. Two Gaussian linear models are compared, one with density and one with resin-adjusted density as variables:
\begin{align}
  \text{Model } M_1 & :          & \datum_i                        & = \alpha + \beta(x_i - \bar{x}) + \epsilon_i ,
                    & \epsilon_i & \sim \text{N}(0, \tau^{-1});                                                     \\
  \text{Model } M_2 & :          & \datum_i                        & = \gamma + \delta(z_i - \bar{z}) + \eta_i ,
                    & \eta_i     & \sim \text{N}(0, \lambda^{-1}),
\end{align}
where $\bar{x}$, $\bar{z}$ denote the mean values of $x_i$ and $z_i$ respectively, and $\tau$ and $\lambda$ denote the precision of the noise for the respective models. For both models, Gaussian priors with means $\mu_\alpha = 3000$ and $\mu_\beta = 185$, and precision scales $r_0 = 0.06$ and $s_0 = 6$ are chosen. A gamma prior is assumed for the noise precision with shape $a_0 = 3$ and rate $b_0 = 2 \times 300^2$.
\begin{figure}
  \centering
  \subfloat[Model 1]{\includegraphics[width=\linewidth]{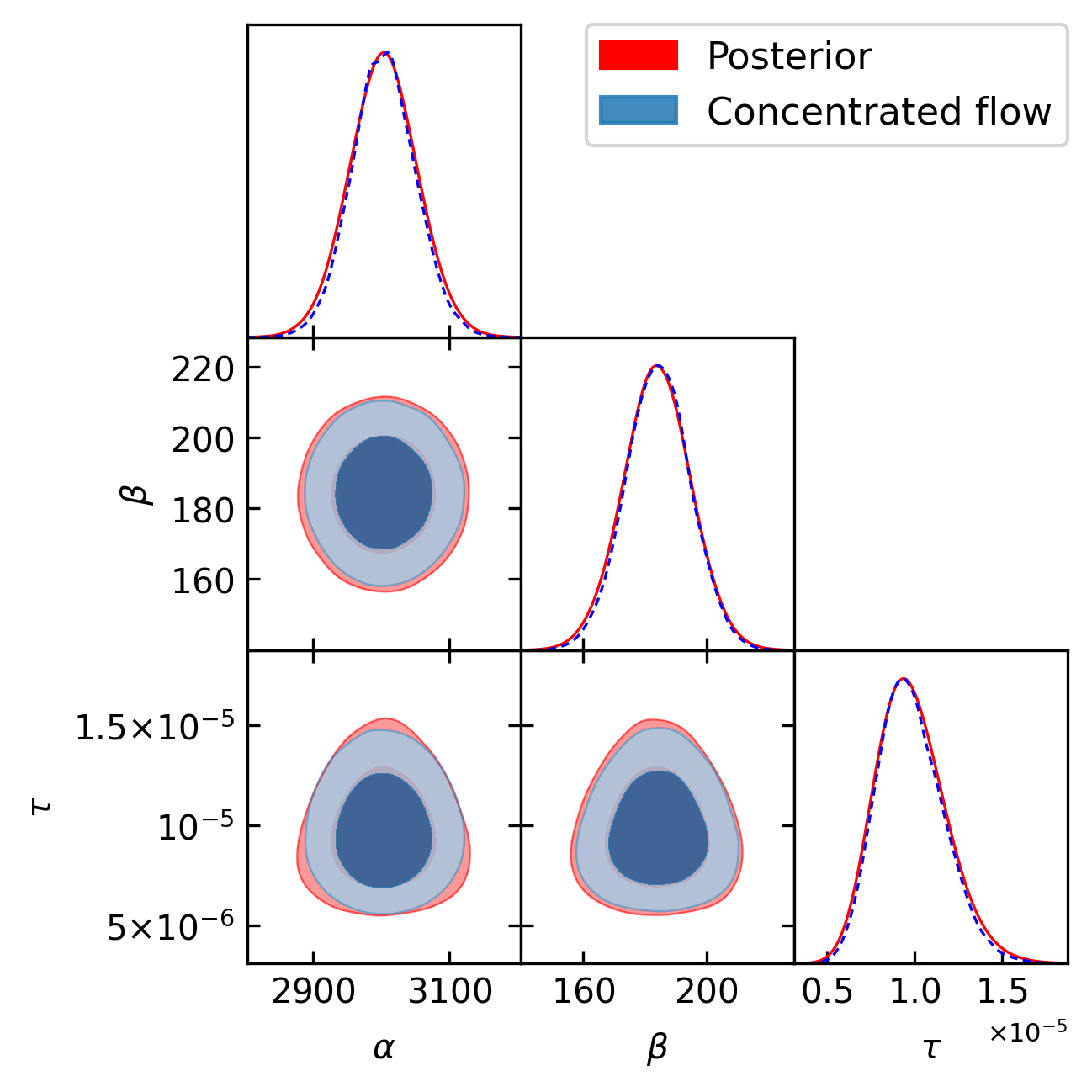}}
  \qquad
  \subfloat[Model 2]{\includegraphics[width=\linewidth]{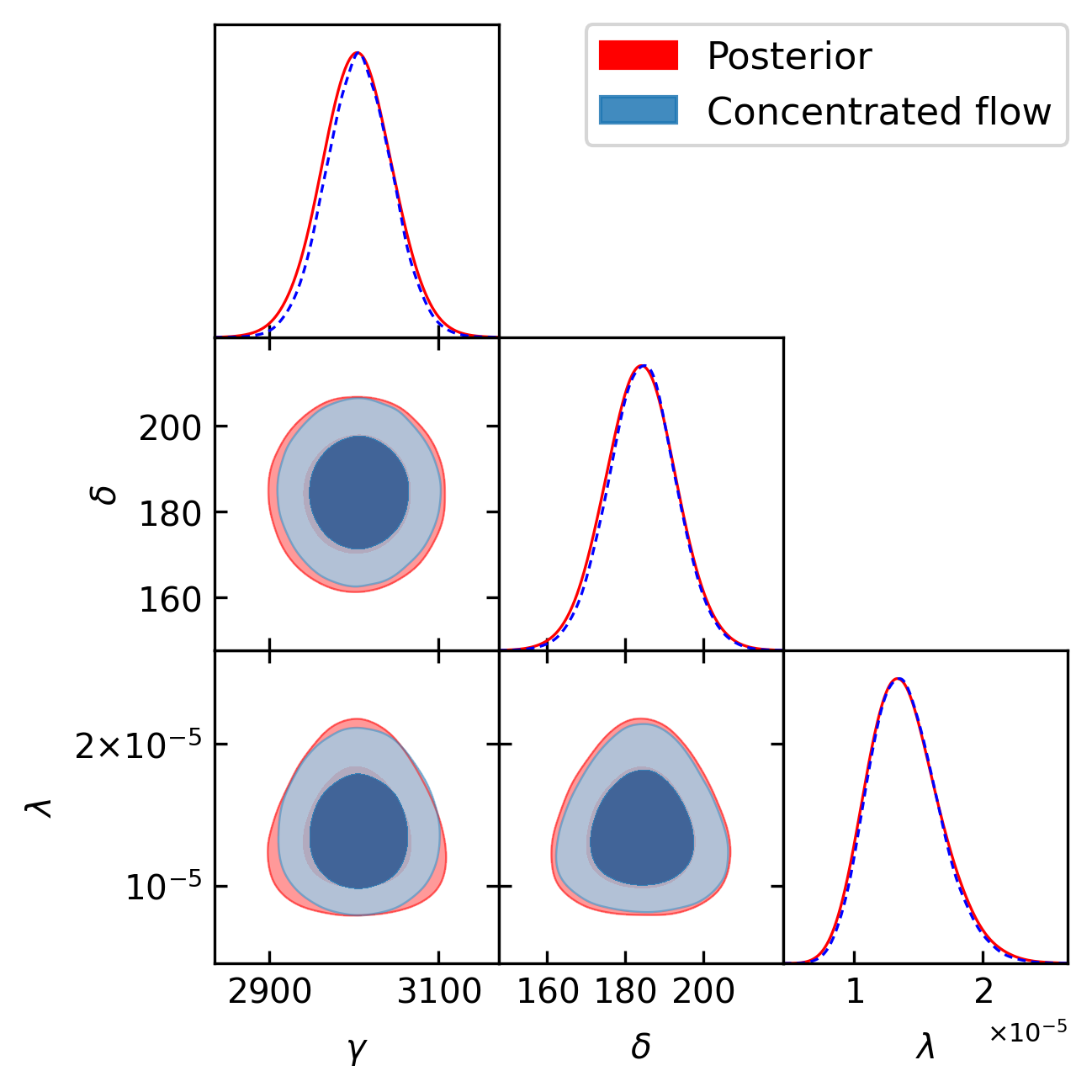}}
  \caption{Corner plot of the the sampled posterior (solid red) and rational quadratic spline flow trained on the posterior samples with temperature $T=0.9$ (dashed blue) for the Radiata pine benchmark problem.
    The internal importance target distribution given by the concentrated flow is contained within the posterior and has thinner tails, as required for the learned harmonic mean estimator.}
  \label{fig:radiata_pine_corner}
\end{figure}
The evidence can be computed analytically for this example \citep{mcewen2023machine}.

Using \texttt{emcee}, we draw $10,000$ samples for $200$ chains, with burn-in of $2,000$ samples, yielding $8,000$ posterior samples per chain. We train a rational quadratic flow consisting of $2$ layers, with $50$ spline bins. Standardization is applied to the data as detailed in Section \ref{sec:learned_harmonic_mean_flows}, which is necessary due to the vast difference in scale of the parameter dimensions.

Figure~\ref{fig:radiata_pine_corner} shows a corner plot of the training samples from the posterior (red) and from the normalizing flow (blue) at temperature $T=0.9$ for both models. Again, it can be seen that the concentrated learned target is contained within the posterior. The log evidence found for Model 1 and 2 is $-310.1284^{0.0007}_{-0.0007}$ and $-301.7044^{0.0008}_{-0.0008}$ respectively, resulting in the estimate $\log\text{BF}_{12}= 8.424^{0.001}_{-0.001}$. The analytic values of the log evidence are $-310.1283$ and $-301.7046$ for Models 1 and 2 respectively, resulting in the estimate $\log\text{BF}_{12}= 8.424$. The value obtained using our estimator is in close agreement with the ground truth. The learned harmonic mean gives an accurate estimate of the evidence, whereas the original harmonic mean estimator fails catastrophically for this example \citep{friel2012estimating}.

\subsection{Gaussian in 21 dimensions}
\begin{figure}
  \centering
  \includegraphics[width=\linewidth]{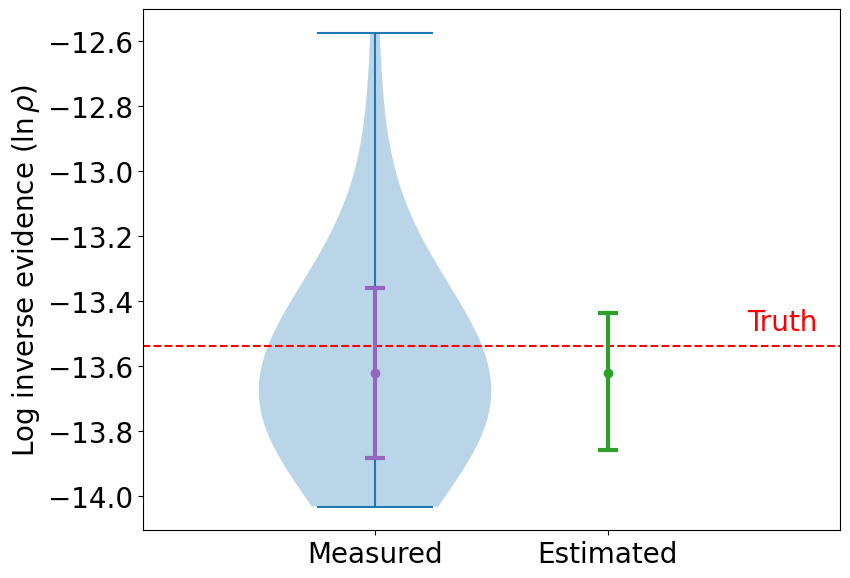}
  \caption{Violin plot of the log reciprocal Bayesian evidence computed by the learned harmonic mean estimator for a 21-dimensional Gaussian benchmark problem repeated $100$ times at
      $T=0.8$. The plot shows log reciprocal Bayesian evidence estimates across runs (measured) along with the one estimate and its error estimate (estimated). The ground truth is shown in red.}
  \label{fig:gaussian_21D}
\end{figure}
To further validate the method and error estimate in a moderate-dimensional context where the ground truth is available, we consider a simple $21$-dimensional Gaussian example. We choose this number of dimensions as it is the same as for the cosmological example we consider in Section~\ref{sec:cosmo}. The diagonal elements are initialised as one plus Gaussian noise with zero mean and unit variance scaled by $0.1$. The off-diagonal elements adjacent to the diagonal are set to be the geometric mean of their adjacent diagonal elements scaled by $0.5$, with alternating signs. We draw $5,000$ MCMC samples for $80$ chains, with burn-in of $500$ samples, yielding $4,500$ posterior samples per chain. We use a rational quadratic flow consisting of $3$ layers, with $128$ spline bins with standardization, at temperature $T=0.8$. We train on half of the chains and use the other half for inference. We repeat this experiment $100$ times with different seeds.

Figure~\ref{fig:gaussian_21D} shows the results of this experiment. The analysis is analogous to the 2-dimensional Rosenbrock experiment described in Section~\ref{sec:rosenbrock}, but in log scale. It can be seen that the estimated value are in agreement with the ground truth, and the estimated and measured errors are similar.
\vspace{0.5cm}

\subsection{Robustness of the temperature parameter}
\label{sec:temperature}

\begin{figure}[t]
  \centering
  \includegraphics[width=1.05\linewidth]{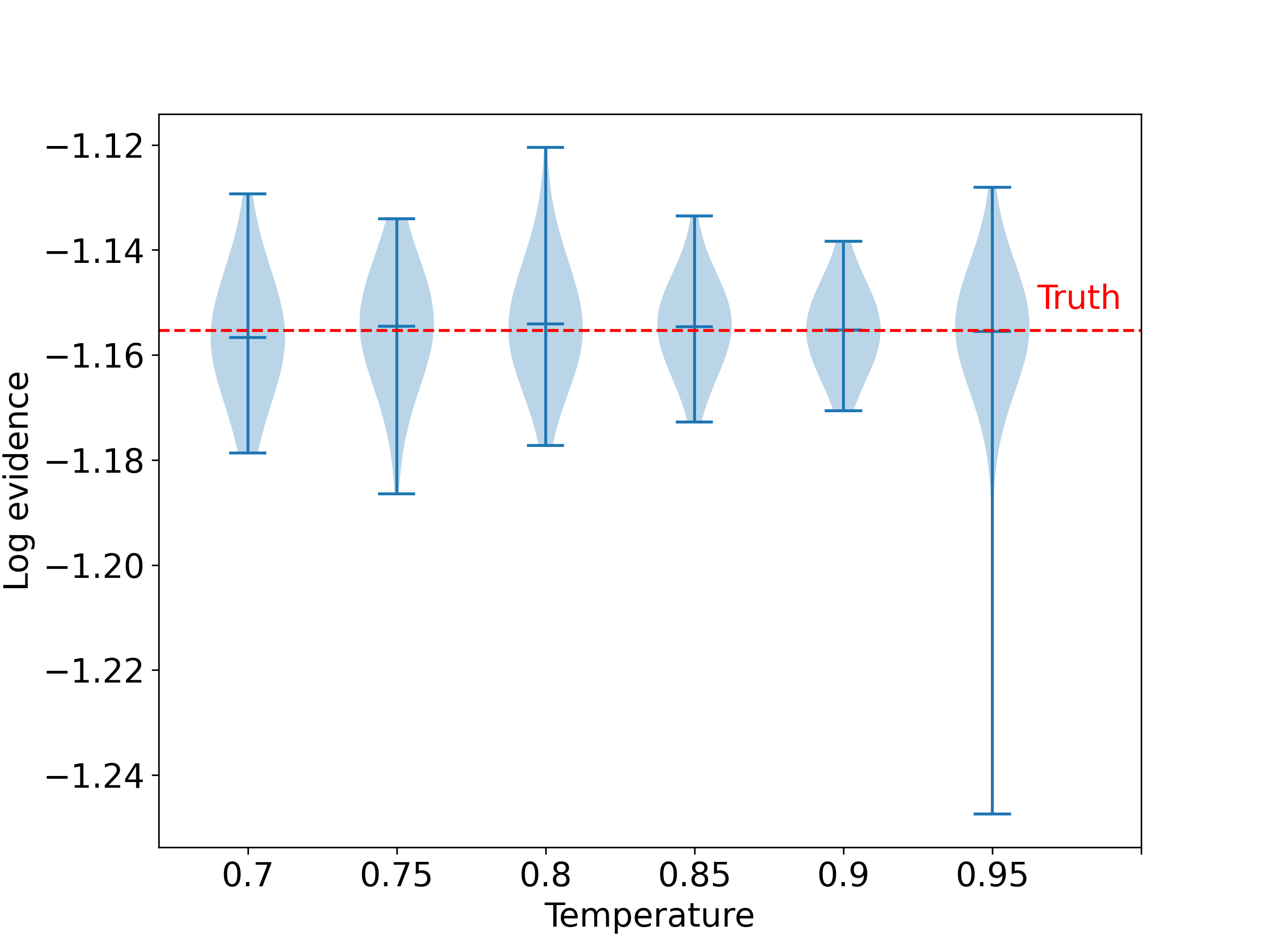}
  \caption{Impact of the temperature parameter value on the evidence estimate. The figure contains violin plots of the evidence estimates across runs for the Rosenbrock problem for a range of temperature values. The ground truth is shown in red. It can be seen that the Bayesian evidence estimates are accurate for a range of temperatures. This shows that the learned harmonic mean is a robust method and does not require careful parameter fine-tuning. The outlier value for $T=0.95$ illustrates the fact that even though the corresponding concentrated flow better approximates the optimal importance target given by the posterior, a flow temperature closer to unity does not necessarily lead to a better estimate since as $T \rightarrow 1$ it is possible the flow may not contain the posterior (as it does not represent the true underlying posterior but only a learned approximation).}
  \label{fig:ev_v_T}
\end{figure}

\begin{figure*}
  \centering
  \subfloat[Log reciprocal evidence for 10D Rosenbrock]{\includegraphics[width=\linewidth]{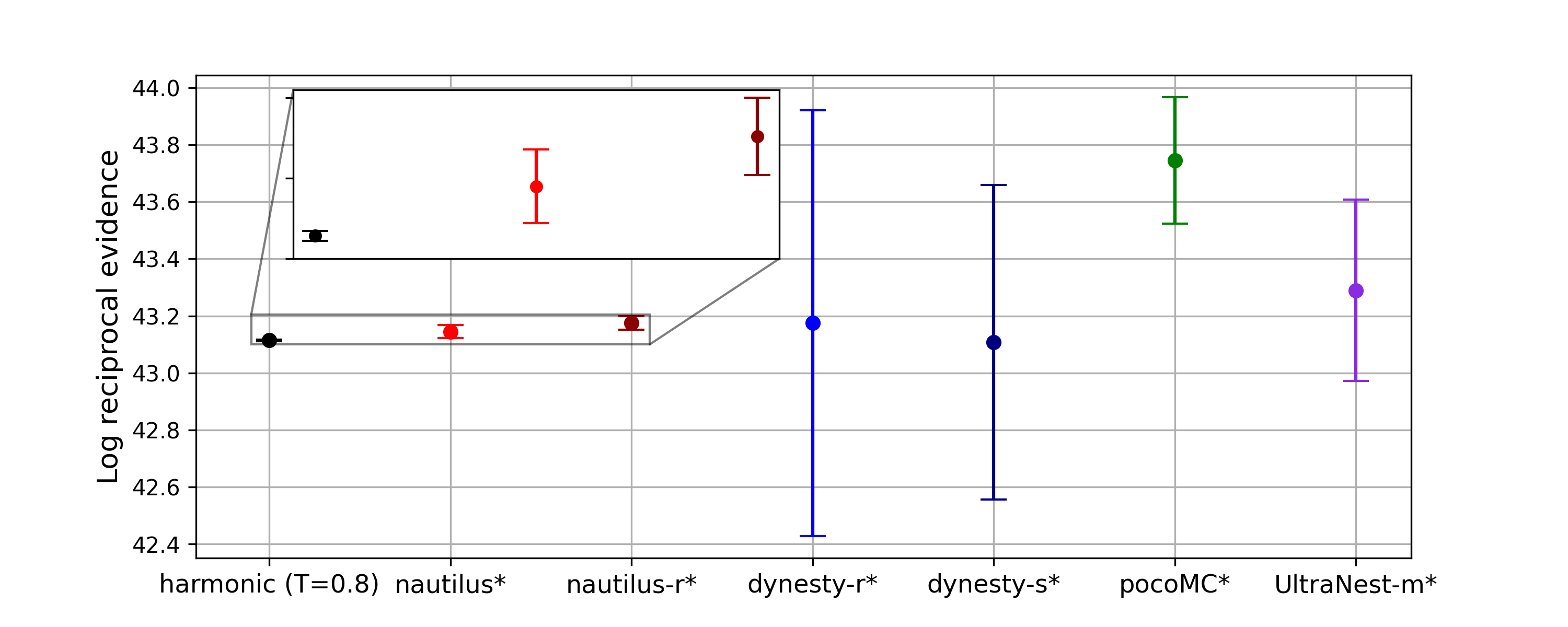} \label{fig:rosenbrock10D_a}}
  \qquad
  \subfloat[Learned harmonic mean log reciprocal evidence for varying temperature]{\includegraphics[width=\linewidth]{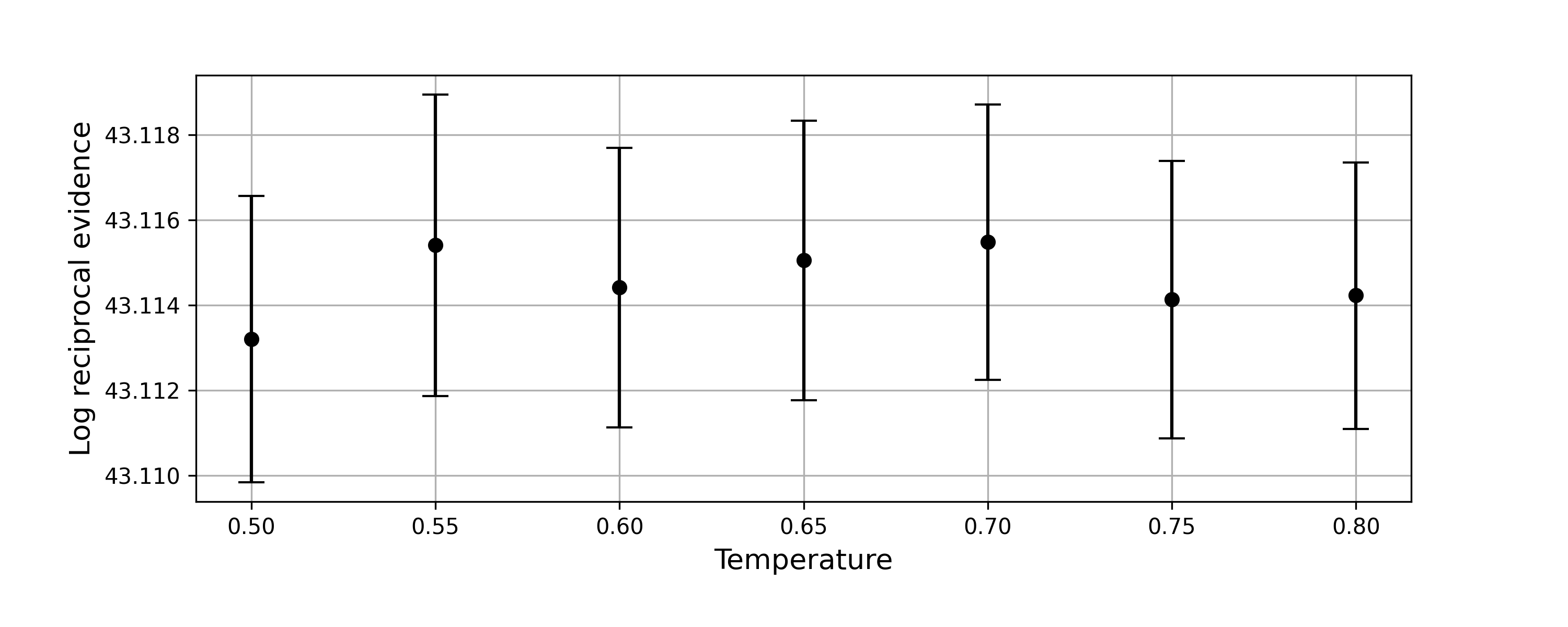} \label{fig:rosenbrock10D_b}}
  \caption{Log reciprocal evidence values for the 10D Rosenbrock example. (a) Comparison of estimates obtained using different methods. Starred methods denote values obtained by \citet{lange2023nautilus}. (b) Estimates obtained using the learned harmonic mean for a range of temperature values. It can be seen that the estimator is robust to this hyperparameter.}
  \label{fig:rosenbrock10D}
\end{figure*}

\begin{table*}
  \caption{Evidence and Bayes factors computed for DES Y1-like 3x2pt analysis}
  \label{table:desy1}
  \centering
  \begin{tabular}{ccccc} \toprule
    Method                 & $\log(z_{\Lambda\text{CDM}})$ & $\log(z_{w\text{CDM}})$    & $\log\text{BF}_{\Lambda\text{CDM-$w$CDM}}$ & Computation time (64 CPU cores for sampling)  \\ \midrule
    Learned harmonic mean  & $-65.262_{-0.011}^{+0.011}$   & $-67.407_{-0.009}^{0.009}$ & $2.145_{-0.014}^{0.014}$                   & 16 hours (sampling) $+$ 16 minutes (evidence) \\
    Nested sampling        & $-65.21 \pm 0.32$             & $-67.44 \pm 0.32$          & $2.23 \pm 0.45$                            & 94 hours (sampling and evidence)              \\
    Na\"ive flow estimator & $-64.9 \pm 0.8$               & $-67.0 \pm 1.1$            & $2.1 \pm 1.4$                              & Similar to learned harmonic mean              \\
    \bottomrule
  \end{tabular}
\end{table*}

Many methods of estimating the evidence require careful fine-tuning of hyperparameters. As explained in Section \ref{sec:learned_harmonic_mean}, this was also the case for the learned harmonic mean estimator when using the classical machine learning models as considered previously. In this work, through the introduction of a more sophisticated machine learning model, normalizing flows, we are able to avoid this drawback and create a more robust estimator.

Our learned harmonic mean estimator with normalizing flows contains essentially just a single hyperparameter: the temperature $T$ of the concentrated flow. We perform numerical experiments to study the influence of the temperature parameter $T$ on the evidence estimate. The Rosenbrock benchmark problem is considered again, as described in Section \ref{sec:rosenbrock}. The experimental process is performed for a range of temperatures $T \in [0.7,0.95]$, repeating it $100$ times for each value. For each repeat, a new seed is used to generate a new dataset of posterior samples, and to initialize the optimizer.

Figure~\ref{fig:ev_v_T} shows violin plots of the log evidence estimates obtained in this experiment plotted for each temperature value. The ground truth, obtained through direct numerical integration is shown in red. It can be seen that the evidence estimates remain accurate and unbiased for the range of temperature values considered. This illustrates the robustness of our method -- the temperature parameter does not need to be fine-tuned.

One must nevertheless ensure that the flow is indeed contained within the posterior as required for the learned harmonic mean to be accurate. The temperature parameter needs to be sufficiently small for this to be the case. If the flow was a perfect approximation of the posterior, any value $T \leq 1$ would do.
In practice this is not the case, and if the temperature is chosen to be too close to unity, the flow might not be contained within the posterior in some regions of the parameter space, causing the estimator's variance to grow.  This effect can partially be seen when looking at the violin plot for $T=0.95$ in Figure~\ref{fig:ev_v_T}. Most of the evidence estimates remain accurate but it can be seen that there exists an outlier. The smallest evidence estimate computed is many standard deviations away from the ground truth. To avoid this, one should always ensure that the flow at the chosen temperature does not have fatter tails than the posterior.
If the flow for $T=1$ were a perfect approximation of the posterior, one would expect the variance of the estimator to increase as $T$ is reduced below unity due to the resulting smaller effective sample size.  However, when dealing with a finite number of samples from the posterior and imperfect approximations, a temperature value closer to unity is not always best. When $T$ is large, the possibility of the flow not being contained within the posterior increases. It is better to choose a lower, more conservative value of $T$ when dealing with a more complicated or high-dimensional posterior. In practice, we find $T\approx0.9$ works well for most problems. A lower $T$ value should be used if the posterior is particularly complex or high-dimensional. This value can then be adjusted based on the error estimate or other diagnostics computed by the \texttt{harmonic} code \citep{mcewen2023machine}.

To investigate the impact of the temperature parameter for a non-Gaussian example of moderate dimensions, we study a 10-dimensional Rosenbrock example considered by \citet{lange2023nautilus}. We consider a range of temperatures $T \in [0.5,0.8]$. We draw $10,000,000$ samples for $200$ chains, with burn-in of $404,800$ samples, yielding $9,595,200$ posterior samples per chain. Note that this is a challenging density to sample in higher dimensions, and \texttt{emcee} is not the optimal choice for this task. For this reason, \citet{lange2023nautilus} consider a much higher number of samples, which we reduce due to computational constraints. Additionally, the aim of this work is not to apply MCMC methods in these challenging settings, but to validate the robustness of the learned harmonic mean for a moderate dimensional non-Gaussian example, hence we choose not to focus our resources on this complicated task. For each temperature, we consider subsets of the sampled chains: we reserve the last 0.2\% of the chains for training, while leaving the rest for inference. We thin the training set by a factor of $5$, and the inference set by a factor of $1,000$. For each subset, we increase the thinning starting index by $10$.

The results of this analysis are shown in Figure~\ref{fig:rosenbrock10D}. Figure~\ref{fig:rosenbrock10D_a} shows the comparison of our log reciprocal evidence estimate at $T=0.8$, alongside the values quoted by \citet{lange2023nautilus}. Our estimate is in broad agreement with many nested sampling methods, while being significantly more precise and using fewer samples.
  Figure~\ref{fig:rosenbrock10D_b} additionally shows the variation of the log reciprocal evidence estimates with temperature: it can be seen that these results are self-consistent, and hence our estimator is robust to this hyperparameter in this setting as well.

\subsection{Practical cosmological example: DES Y1 analysis}
\label{sec:cosmo}

\begin{figure*}
  \centering
  \subfloat[$\Lambda$CDM]{\includegraphics[width=0.6\linewidth]{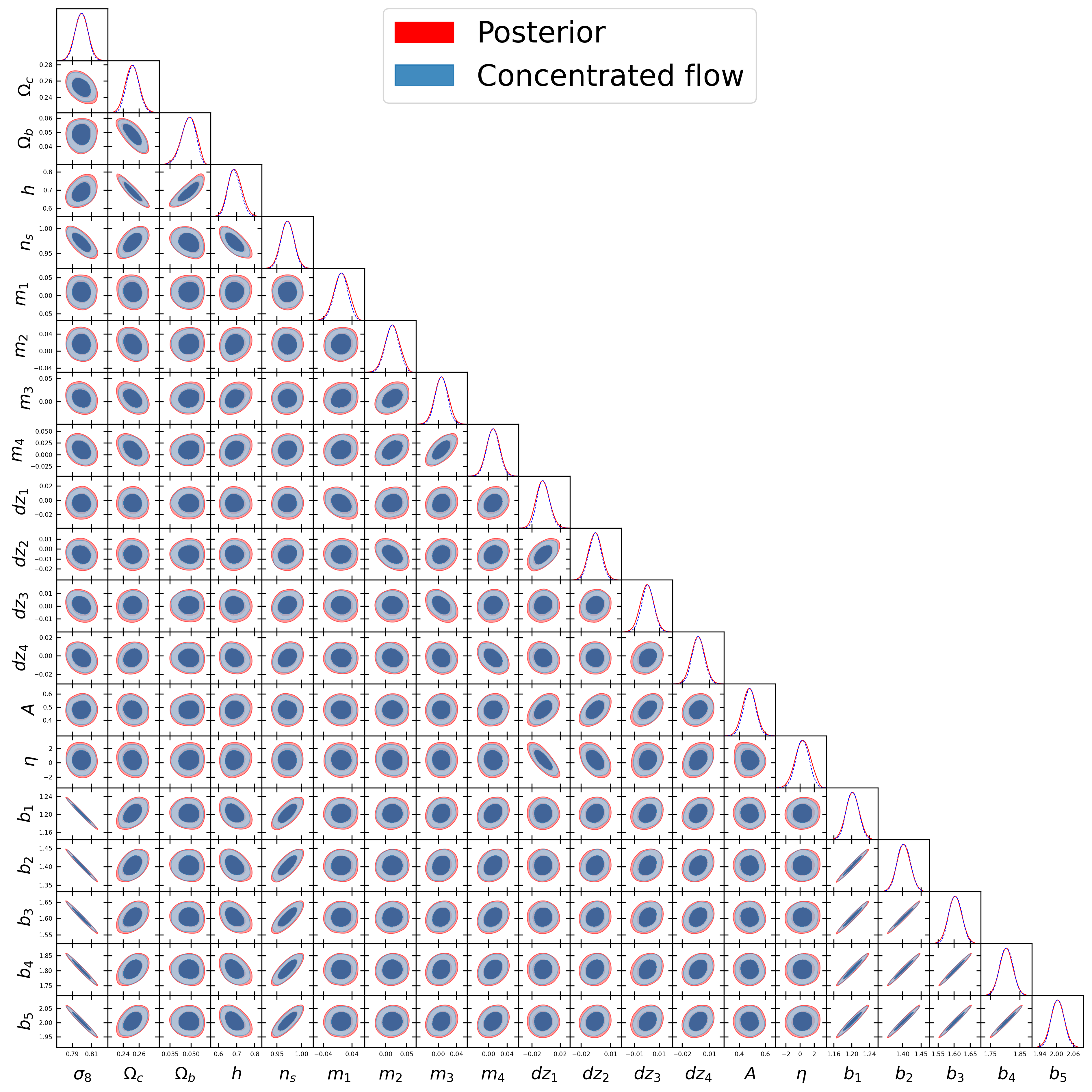}}
  \qquad
  \subfloat[$w$CDM]{\includegraphics[width=0.6\linewidth]{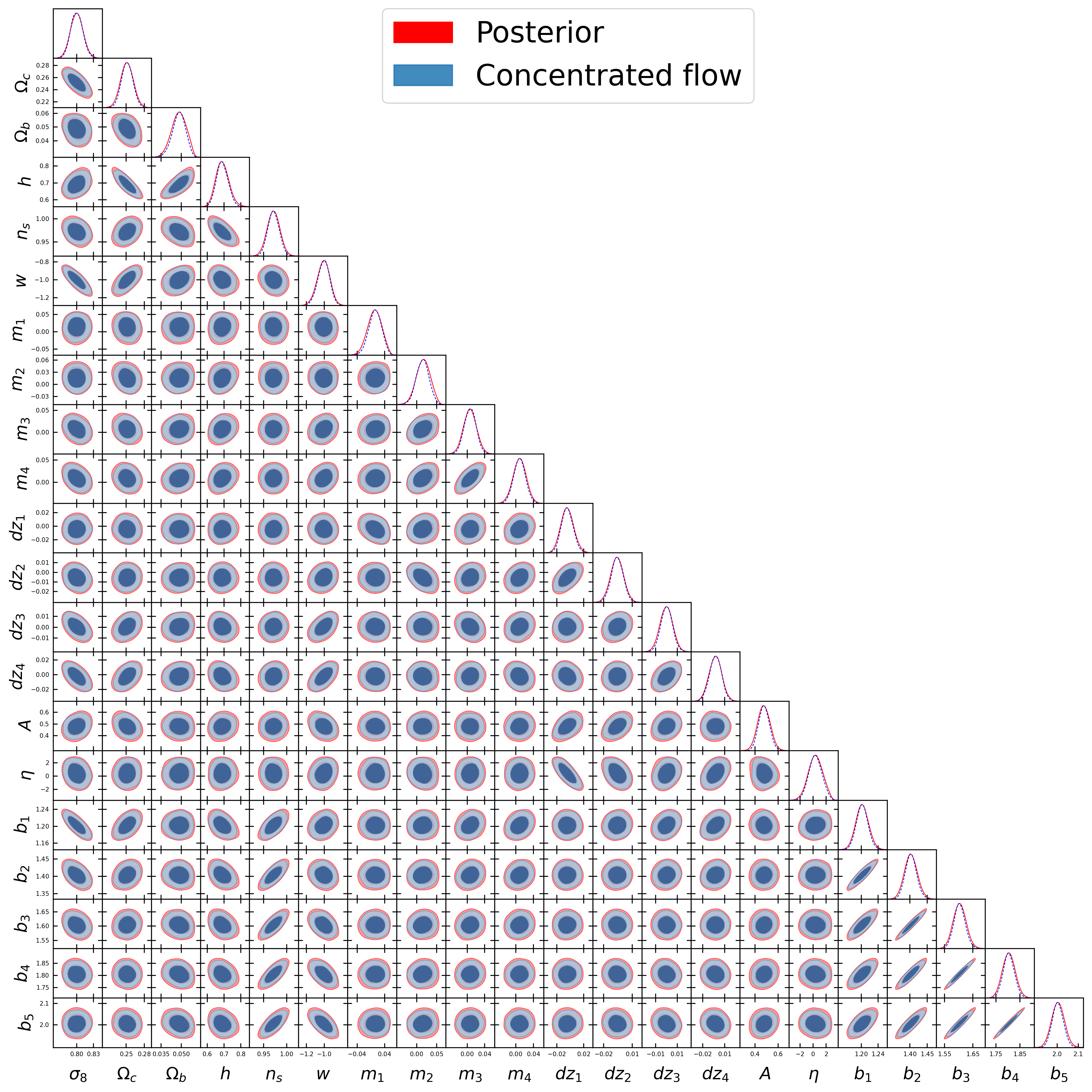}}
  \caption{Corner plot of the sampled posterior (solid red) and rational quadratic spline flow trained on the posterior samples with temperature $T=0.8$ (dashed blue) for the DES Y1 analysis example for (a) $\Lambda$CDM and (b) $w$CDM cosmological models.
    The internal importance target distribution given by the concentrated flow is contained within the posterior and has thinner tails, as required for the learned harmonic mean estimator, even in this higher dimensional case.}
  \label{fig:cosmo}
\end{figure*}

In Section \ref{sec:benchmarks} we showed that the learned harmonic mean estimator with normalizing flows works very well on a range of simple benchmark examples, where the ground truth Bayesian evidence is available.  In this section we show it also performs well in a practical context, by applying it to a Dark Energy Survey Year 1 (DES Y1) example. DES is an on-going cosmological survey designed to gain insight into the nature of dark energy. We perform a 3x2pt analysis \citep{Abbott_2018,joachimi_2010}, i.e.\ a joint analysis of galaxy clustering and weak lensing considering shear, clustering and their cross correlation, on a DES Y1-like configuration.

We follow the reference approach described by \citet{Campagne_2023}, who extract and compress a subset of the DES Year 1 lensing and clustering data and set up a forward model following the DES Y1 Pipeline \citep{Abbott_2018}. We use the DES Y1 redshift distributions\footnote{\url{http://desdr-server.ncsa.illinois.edu/despublic/y1a1_files/chains/2pt_NG_mcal_1110.fits}} and simulate a 3x2pt data vector for a fixed cosmology. We use this as our mock data vector and run an inference pipeline to obtain posterior contours and evidence estimates. We refer the reader to \citet{Campagne_2023} for the priors and all other details. To sample from the posterior we use the \texttt{Cobaya} package \citep{torrado2019} with the Metropolis-Hastings algorithm. We then apply \texttt{harmonic} to these samples to evaluate the Bayesian evidence. For comparison, we also sample using the \texttt{PolyChord} nested sampler \citep{Handley_2015a,Handley_2015} in \texttt{Cobaya}, which provides a benchmark Bayesian evidence estimate. We perform the analysis twice, assuming either a $\Lambda$CDM or $w$CDM cosmological model, with the dark energy equation of state parameter $w$ fixed to $w=-1$ or free to vary, respectively.

The $\Lambda$CDM and $w$CDM models have $20$ and $21$ parameters respectively. With the Metropolis-Hastings sampler, we run $64$ chains per model, obtaining an average of approximately $5,800$ and $6,500$ samples per chain for $\Lambda$CDM and $w$CDM respectively. We discard $500$ samples for burn-in in both cases. We train a rational quadratic flow consisting of $3$ layers, with $128$ spline bins, applying standardization, on half of the chains, and use the other half for inference.

Figure~\ref{fig:cosmo} shows the corner plots of the training samples alongside the flow at
$T=0.8$. It can be seen the flow also behaves as expected in this higher-dimensional practical setting, capturing the posterior distribution while being contained within it. Log evidence values, Bayes factors and computation time are reported in Table~\ref{table:desy1} for the learned harmonic mean estimator, nested sampling with \texttt{PolyChord} and by the na\"ive flow estimate introduced in \citet{Spurio_Mancini_2023} and described in Section~\ref{sec:naive_estimate}.

Note that the values computed by the learned harmonic mean and nested sampling are in agreement, showing a slight preference for $\Lambda$CDM, matching the configuration of our simulated setup. The values computed by the na\"ive estimator are in approximate agreement but exhibit an error two orders or magnitude larger than the error of the learned harmonic mean (in higher dimensional examples to be reported in an ongoing work we observe the na\"ive estimator failing much more catastrophically).

In terms of computational speed (summarized in Table~\ref{table:desy1} but reported in great detail here), sampling with \texttt{Cobaya} using the Metropolis-Hastings algorithm takes approximately $8$ hours for $\Lambda$CDM and $w$CDM each on $64$ CPU cores. The compute time added by \texttt{harmonic} is around $5$ minutes on $1$ GPU for training and $3$ minutes on $128$ CPU cores to estimate the evidence for each model. Using \texttt{PolyChord} takes approximately $47$ hours for $\Lambda$CDM and $w$CDM each, on the same $64$ CPU cores used for the Metropolis-Hastings sampling. The Metropolis-Hastings algorithm in this case is much quicker due to the use of a proposal covariance matrix based on a Planck cosmology \citep{Campagne_2023}. Thanks to the flexibility of the learned harmonic estimator, we can leverage this advantage and choose Metropolis-Hastings over nested sampling, while still being able to estimate the Bayesian evidence and perform model comparison. Even in this higher dimensional setting, the learned harmonic mean only adds a few minutes of compute time on top of the sampler. This demonstrates the potential scalability of the method and its potential for computing the evidence from existing saved down MCMC chains.

\section{Conclusions}
\label{sec:conclusions}

In this work we outlined the learned harmonic mean estimator with normalizing flows, a robust, flexible and scalable estimator of the Bayesian evidence.  Normalizing flows meet the core requirements of the learned importance target distribution of the learned harmonic mean estimator: namely, they provide a normalized probability distribution for which one can evaluate probability densities. We use them to introduce an elegant way to ensure the probability mass of the learned distribution is contained within the posterior, a critical requirement of the learned harmonic mean.  This avoids the need for a bespoke training approach, resulting in a more robust and flexible estimator.  Furthermore, flows offer the potential of greater scalability than the classical machine learning models considered previously.

To validate its accuracy, we applied the learned harmonic mean to several benchmark problems. Our method produced accurate results, even in cases where the original harmonic mean had been shown to fail. We also applied the learned harmonic mean to a practical cosmological example, the Dark Energy Survey Year 1 (DES Y1) data 3x2pt analysis. Even in this higher dimensional context for up to $21$ parameters our method computed an estimate that was in excellent agreement with the conventional approach using nested sampling. This shows the potential for scalability of our method.  Many existing methods of estimating the Bayesian evidence, including previous work on the learned harmonic mean, require careful parameter fine-tuning. Beyond the flow architecture, we only introduced one hyperparameter -- the concentrated flow temperature $T$, which does not require any fine-tuning. We showed this empirically by considering a selected benchmark problem for a range of $T$ values. The estimate remained accurate, demonstrating the robustness of our method.

Since the learned harmonic mean estimator is decoupled from the sampling method, it can be used in a wide variety of settings. This includes approaches such as simulation-based inference, variational inference and various MCMC methods where the evidence could not otherwise be computed accurately, such as the No U-Turn Sampler (NUTS) \citep{hoffman2011nouturn}. When using MCMC methods for parameter estimation, the Bayesian evidence can be obtained essentially ``for free'' or even post-hoc from saved down MCMC chains.  Since the estimator is agnostic to the sampling strategy, it is highly flexible. The best suited sampling strategy may be used for the problem at hand, as we demonstrated in the DES Y1 example, where Metropolis-Hastings accurately sampled the posterior much faster than nested sampling. In recent work we leverage the flexibility of the learned harmonic mean to demonstrate its use with NUTS and the CosmoPower-JAX emulator \citep{spurio2022cosmopower, Piras_2023} to scale evidence calculation to $\sim 150$ dimensions \citep{piras2024future}.  Overall, the learned harmonic mean estimator with normalizing flows is a robust, flexible and scalable tool for Bayesian model comparison that can be used in a variety of contexts.

\section*{Acknowledgements}
We thank Kaze Wong for insightful discussions regarding normalizing flows. A.P.\ is supported by the UCL Centre for Doctoral Training in Data Intensive Science (STFC grant number ST/W00674X/1). M.A.P.\ and J.D.M.\ are supported by EPSRC (grant number EP/W007673/1). D.P.\ was supported by a Swiss National Science Foundation (SNSF) Professorship grant (No. 202671), and by the SNF Sinergia grant CRSII5-193826 ``AstroSignals: A New Window on the Universe, with the New Generation of Large Radio-Astronomy Facilities''. A.S.M.\ acknowledges support from the MSSL STFC Consolidated Grant ST/W001136/1.

\bibliographystyle{mnras}
\bibliography{references}

\begin{thebibliography}{}
\makeatletter
\relax
\def\mn@urlcharsother{\let\do\@makeother \do\$\do\&\do\#\do\^\do\_\do\%\do\~}
\def\mn@doi{\begingroup\mn@urlcharsother \@ifnextchar [ {\mn@doi@} {\mn@doi@[]}}
\def\mn@doi@[#1]#2{\def\@tempa{#1}\ifx\@tempa\@empty \href {http://dx.doi.org/#2} {doi:#2}\else \href {http://dx.doi.org/#2} {#1}\fi \endgroup}
\def\mn@eprint#1#2{\mn@eprint@#1:#2::\@nil}
\def\mn@eprint@arXiv#1{\href {http://arxiv.org/abs/#1} {{\tt arXiv:#1}}}
\def\mn@eprint@dblp#1{\href {http://dblp.uni-trier.de/rec/bibtex/#1.xml} {dblp:#1}}
\def\mn@eprint@#1:#2:#3:#4\@nil{\def\@tempa {#1}\def\@tempb {#2}\def\@tempc {#3}\ifx \@tempc \@empty \let \@tempc \@tempb \let \@tempb \@tempa \fi \ifx \@tempb \@empty \def\@tempb {arXiv}\fi \@ifundefined {mn@eprint@\@tempb}{\@tempb:\@tempc}{\expandafter \expandafter \csname mn@eprint@\@tempb\endcsname \expandafter{\@tempc}}}

\bibitem[\protect\citeauthoryear{Abbott et~al.,}{Abbott et~al.}{2018a}]{Abbott_2018}
Abbott T.,  et~al., 2018a, \mn@doi [Physical Review D] {10.1103/physrevd.98.043526}, 98

\bibitem[\protect\citeauthoryear{Abbott et~al.,}{Abbott et~al.}{2018b}]{abbott2018dark}
Abbott T.~M.,  et~al., 2018b, Physical Review D, 98, 043526

\bibitem[\protect\citeauthoryear{Aghanim et~al.,}{Aghanim et~al.}{2020}]{aghanim2020planck}
Aghanim N.,  et~al., 2020, Astronomy \& Astrophysics, 641, A6

\bibitem[\protect\citeauthoryear{Ashton et~al.,}{Ashton et~al.}{2022}]{ashton2022nested}
Ashton G.,  et~al., 2022, Nature Reviews Methods Primers, 2, 39

\bibitem[\protect\citeauthoryear{Bernardo \& Smith}{Bernardo \& Smith}{1994}]{bernardo:1994}
Bernardo J.,  Smith A.,  1994, \mn@doi [Wiley Online Library] {https://doi.org/10.1002/9780470316870.ch5}

\bibitem[\protect\citeauthoryear{Bradbury et~al.,}{Bradbury et~al.}{2018}]{jax2018github}
Bradbury J.,  et~al., 2018, {JAX}: composable transformations of {P}ython+{N}um{P}y programs, \url {http://github.com/google/jax}

\bibitem[\protect\citeauthoryear{Brewer, P{\'a}rtay  \& Cs{\'a}nyi}{Brewer et~al.}{2011}]{2010ascl.soft10029B}
Brewer B.~J.,  P{\'a}rtay L.~B.,   Cs{\'a}nyi G.,  2011, Statistics and Computing, 21, 649

\bibitem[\protect\citeauthoryear{Brout et~al.,}{Brout et~al.}{2022}]{Brout_2022}
Brout D.,  et~al., 2022, \mn@doi [The Astrophysical Journal] {10.3847/1538-4357/ac8e04}, 938, 110

\bibitem[\protect\citeauthoryear{Buchner}{Buchner}{2021}]{Buchner2021}
Buchner J.,  2021, \mn@doi [Journal of Open Source Software] {10.21105/joss.03001}, 6, 3001

\bibitem[\protect\citeauthoryear{Cai, McEwen  \& Pereyra}{Cai et~al.}{2022}]{cai2022proximal}
Cai X.,  McEwen J.~D.,   Pereyra M.,  2022, Statistics and Computing, 32

\bibitem[\protect\citeauthoryear{Campagne et~al.,}{Campagne et~al.}{2023}]{Campagne_2023}
Campagne J.-E.,  et~al., 2023, \mn@doi [The Open Journal of Astrophysics] {10.21105/astro.2302.05163}, 6

\bibitem[\protect\citeauthoryear{Carron, Mirmelstein  \& Lewis}{Carron et~al.}{2022}]{carron2022cmb}
Carron J.,  Mirmelstein M.,   Lewis A.,  2022, Journal of Cosmology and Astroparticle Physics, 2022, 039

\bibitem[\protect\citeauthoryear{Chib}{Chib}{1995}]{chib:1995}
Chib S.,  1995, Journal of the American Statistical Association, 90, 1313

\bibitem[\protect\citeauthoryear{Clyde, Berger, Bullard, Ford, Jefferys, Luo, Paulo  \& Loredo}{Clyde et~al.}{2007}]{clyde2007current}
Clyde M.,  Berger J.,  Bullard F.,  Ford E.,  Jefferys W.,  Luo R.,  Paulo R.,   Loredo T.,  2007, in Statistical challenges in modern astronomy IV. p.~224

\bibitem[\protect\citeauthoryear{{DES Collaboration} et~al.,}{{DES Collaboration} et~al.}{2024}]{descollaboration2024dark}
{DES Collaboration} et~al., 2024, The Dark Energy Survey: Cosmology Results With ~1500 New High-redshift Type Ia Supernovae Using The Full 5-year Dataset (\mn@eprint {arXiv} {2401.02929})

\bibitem[\protect\citeauthoryear{{DESI Collaboration} et~al.,}{{DESI Collaboration} et~al.}{2016}]{desicollaboration2016desi}
{DESI Collaboration} et~al., 2016, The DESI Experiment Part I: Science,Targeting, and Survey Design (\mn@eprint {arXiv} {1611.00036})

\bibitem[\protect\citeauthoryear{DeepMind et~al.,}{DeepMind et~al.}{2020}]{deepmind2020jax}
DeepMind et~al., 2020, The {D}eep{M}ind {JAX} {E}cosystem, \url {http://github.com/deepmind}

\bibitem[\protect\citeauthoryear{Dillon et~al.,}{Dillon et~al.}{2017}]{dillon2017tensorflow}
Dillon J.~V.,  et~al., 2017, TensorFlow Distributions (\mn@eprint {arXiv} {1711.10604})

\bibitem[\protect\citeauthoryear{Dinh, Sohl-Dickstein  \& Bengio}{Dinh et~al.}{2017}]{dinh2016density}
Dinh L.,  Sohl-Dickstein J.,   Bengio S.,  2017, in International Conference on Learning Representations. \url {https://openreview.net/forum?id=HkpbnH9lx}

\bibitem[\protect\citeauthoryear{Dozat}{Dozat}{2016}]{dozat.2016}
Dozat T.,  2016, in Proceedings of the 4th International Conference on Learning Representations. pp~1--4

\bibitem[\protect\citeauthoryear{Durkan, Bekasov, Murray  \& Papamakarios}{Durkan et~al.}{2019}]{durkan2019neural}
Durkan C.,  Bekasov A.,  Murray I.,   Papamakarios G.,  2019, Advances in neural information processing systems, 32

\bibitem[\protect\citeauthoryear{Feroz \& Hobson}{Feroz \& Hobson}{2008}]{Feroz_2008}
Feroz F.,  Hobson M.~P.,  2008, \mn@doi [Monthly Notices of the Royal Astronomical Society] {10.1111/j.1365-2966.2007.12353.x}, 384, 449–463

\bibitem[\protect\citeauthoryear{{Feroz}, {Hobson}  \& {Bridges}}{{Feroz} et~al.}{2009a}]{2009MNRAS.398.1601F}
{Feroz} F.,  {Hobson} M.~P.,   {Bridges} M.,  2009a, \mn@doi [\mnras] {10.1111/j.1365-2966.2009.14548.x}, \href {https://ui.adsabs.harvard.edu/abs/2009MNRAS.398.1601F} {398, 1601}

\bibitem[\protect\citeauthoryear{Feroz, Hobson  \& Bridges}{Feroz et~al.}{2009b}]{Feroz_2009}
Feroz F.,  Hobson M.~P.,   Bridges M.,  2009b, \mn@doi [Monthly Notices of the Royal Astronomical Society] {10.1111/j.1365-2966.2009.14548.x}, 398, 1601–1614

\bibitem[\protect\citeauthoryear{Feroz, Hobson, Cameron  \& Pettitt}{Feroz et~al.}{2019}]{Feroz_2019}
Feroz F.,  Hobson M.~P.,  Cameron E.,   Pettitt A.~N.,  2019, \mn@doi [The Open Journal of Astrophysics] {10.21105/astro.1306.2144}, 2

\bibitem[\protect\citeauthoryear{{Foreman-Mackey}, {Hogg}, {Lang}  \& {Goodman}}{{Foreman-Mackey} et~al.}{2013}]{emcee}
{Foreman-Mackey} D.,  {Hogg} D.~W.,  {Lang} D.,   {Goodman} J.,  2013, PASP, 125, 306

\bibitem[\protect\citeauthoryear{Friel \& Wyse}{Friel \& Wyse}{2012}]{friel2012estimating}
Friel N.,  Wyse J.,  2012, Statistica Neerlandica, 66, 288

\bibitem[\protect\citeauthoryear{Gelfand \& Dey}{Gelfand \& Dey}{1994}]{gelfand1994bayesian}
Gelfand A.~E.,  Dey D.~K.,  1994, Journal of the Royal Statistical Society: Series B (Methodological), 56, 501

\bibitem[\protect\citeauthoryear{Green}{Green}{1995}]{green:1995}
Green P.~J.,  1995, Biometrika, 82, 711

\bibitem[\protect\citeauthoryear{Handley, Hobson  \& Lasenby}{Handley et~al.}{2015a}]{Handley_2015a}
Handley W.~J.,  Hobson M.~P.,   Lasenby A.~N.,  2015a, \mn@doi [Monthly Notices of the Royal Astronomical Society: Letters] {10.1093/mnrasl/slv047}, 450, L61–L65

\bibitem[\protect\citeauthoryear{Handley, Hobson  \& Lasenby}{Handley et~al.}{2015b}]{Handley_2015}
Handley W.~J.,  Hobson M.~P.,   Lasenby A.~N.,  2015b, Monthly Notices of the Royal Astronomical Society, 453, 4385–4399

\bibitem[\protect\citeauthoryear{Hastings}{Hastings}{1970}]{hastings:1970}
Hastings W.~K.,  1970, Biometrika, 57, 97

\bibitem[\protect\citeauthoryear{Heavens, Fantaye, Mootoovaloo, Eggers, Hosenie, Kroon  \& Sellentin}{Heavens et~al.}{2017}]{heavens2017marginal}
Heavens A.,  Fantaye Y.,  Mootoovaloo A.,  Eggers H.,  Hosenie Z.,  Kroon S.,   Sellentin E.,  2017, Marginal Likelihoods from Monte Carlo Markov Chains (\mn@eprint {arXiv} {1704.03472})

\bibitem[\protect\citeauthoryear{Heek, Levskaya, Oliver, Ritter, Rondepierre, Steiner  \& van {Z}ee}{Heek et~al.}{2023}]{flax2020github}
Heek J.,  Levskaya A.,  Oliver A.,  Ritter M.,  Rondepierre B.,  Steiner A.,   van {Z}ee M.,  2023, {F}lax: A neural network library and ecosystem for {JAX}, \url {http://github.com/google/flax}

\bibitem[\protect\citeauthoryear{Hoffman \& Gelman}{Hoffman \& Gelman}{2011}]{hoffman2011nouturn}
Hoffman M.~D.,  Gelman A.,  2011, The No-U-Turn Sampler: Adaptively Setting Path Lengths in Hamiltonian Monte Carlo (\mn@eprint {arXiv} {1111.4246})

\bibitem[\protect\citeauthoryear{Jia \& Seljak}{Jia \& Seljak}{2020}]{pmlr-v118-jia20a}
Jia H.,  Seljak U.,  2020, in Zhang C.,  Ruiz F.,  Bui T.,  Dieng A.~B.,   Liang D.,  eds,  Proceedings of Machine Learning Research Vol. 118, Proceedings of The 2nd Symposium on Advances in Approximate Bayesian Inference. PMLR, pp 1--14, \url {https://proceedings.mlr.press/v118/jia20a.html}

\bibitem[\protect\citeauthoryear{Joachimi \& Bridle}{Joachimi \& Bridle}{2010}]{joachimi_2010}
Joachimi B.,  Bridle S.~L.,  2010, \mn@doi [A&A] {10.1051/0004-6361/200913657}, 523, A1

\bibitem[\protect\citeauthoryear{Kingma \& Ba}{Kingma \& Ba}{2017}]{kingma2017adam}
Kingma D.~P.,  Ba J.,  2017, Adam: A Method for Stochastic Optimization (\mn@eprint {arXiv} {1412.6980})

\bibitem[\protect\citeauthoryear{Lange}{Lange}{2023}]{lange2023nautilus}
Lange J.~U.,  2023, Monthly Notices of the Royal Astronomical Society, 525, 3181

\bibitem[\protect\citeauthoryear{Lenk}{Lenk}{2009}]{lenk:2009}
Lenk P.,  2009, Journal of Computational and Graphical Statistics, 18, 941

\bibitem[\protect\citeauthoryear{Madhavacheril et~al.,}{Madhavacheril et~al.}{2024}]{madhavacheril2024atacama}
Madhavacheril M.~S.,  et~al., 2024, The Astrophysical Journal, 962, 113

\bibitem[\protect\citeauthoryear{McEwen, Wallis, Price  \& Mancini}{McEwen et~al.}{2021}]{mcewen2023machine}
McEwen J.~D.,  Wallis C. G.~R.,  Price M.~A.,   Mancini A.~S.,  2021, Machine learning assisted Bayesian model comparison: learnt harmonic mean estimator (\mn@eprint {arXiv} {2111.12720})

\bibitem[\protect\citeauthoryear{Metropolis, Rosenbluth, Rosenbluth, Teller  \& Teller}{Metropolis et~al.}{1953}]{metropolis:1953}
Metropolis N.,  Rosenbluth A.~W.,  Rosenbluth M.~N.,  Teller A.~H.,   Teller E.,  1953, J. Chemical Physics, 21, 1087

\bibitem[\protect\citeauthoryear{Murphy}{Murphy}{2012}]{pml1Book}
Murphy K.~P.,  2012, Machine Learning: A Probabilistic Perspective.
The MIT Press, \url {https://probml.github.io/pml-book/}

\bibitem[\protect\citeauthoryear{Neal}{Neal}{1994}]{neal:1994}
Neal R.~M.,  1994, JR Stat Soc Ser A (Methodological), 56, 41

\bibitem[\protect\citeauthoryear{Newton \& Raftery}{Newton \& Raftery}{1994}]{newton1994approximate}
Newton M.~A.,  Raftery A.~E.,  1994, Journal of the Royal Statistical Society: Series B (Methodological), 56, 3

\bibitem[\protect\citeauthoryear{Papamakarios, Nalisnick, Rezende, Mohamed  \& Lakshminarayanan}{Papamakarios et~al.}{2021}]{papamakarios2021normalizing}
Papamakarios G.,  Nalisnick E.,  Rezende D.~J.,  Mohamed S.,   Lakshminarayanan B.,  2021, The Journal of Machine Learning Research, 22, 2617

\bibitem[\protect\citeauthoryear{Piras \& Spurio~Mancini}{Piras \& Spurio~Mancini}{2023}]{Piras_2023}
Piras D.,  Spurio~Mancini A.,  2023, The Open Journal of Astrophysics, 6

\bibitem[\protect\citeauthoryear{Piras, Polanska, Mancini, Price  \& McEwen}{Piras et~al.}{2024}]{piras2024future}
Piras D.,  Polanska A.,  Mancini A.~S.,  Price M.~A.,   McEwen J.~D.,  2024, The future of cosmological likelihood-based inference: accelerated high-dimensional parameter estimation and model comparison (\mn@eprint {arXiv} {2405.12965})

\bibitem[\protect\citeauthoryear{Polanska, Price, Spurio~Mancini  \& McEwen}{Polanska et~al.}{2023}]{polanska2023learned}
Polanska A.,  Price M.~A.,  Spurio~Mancini A.,   McEwen J.~D.,  2023, Physical Sciences Forum, 9

\bibitem[\protect\citeauthoryear{Raftery, Newton, Satagopan  \& Krivitsky}{Raftery et~al.}{2006}]{raftery:2006}
Raftery A.~E.,  Newton M.~A.,  Satagopan J.~M.,   Krivitsky P.~N.,  2006, Preprint

\bibitem[\protect\citeauthoryear{Robert \& Wraith}{Robert \& Wraith}{2009}]{robert:2009}
Robert C.~P.,  Wraith D.,  2009, in Aip conference proceedings. pp 251--262

\bibitem[\protect\citeauthoryear{Rubin et~al.,}{Rubin et~al.}{2023}]{rubin2023union}
Rubin D.,  et~al., 2023, Union Through UNITY: Cosmology with 2,000 SNe Using a Unified Bayesian Framework (\mn@eprint {arXiv} {2311.12098})

\bibitem[\protect\citeauthoryear{Skilling}{Skilling}{2006}]{skilling2006nested}
Skilling J.,  2006, Bayesian Analysis, 1, 833

\bibitem[\protect\citeauthoryear{Smith, Everhart, Dickson, Knowler  \& Johannes}{Smith et~al.}{1988}]{smith1988using}
Smith J.~W.,  Everhart J.~E.,  Dickson W.,  Knowler W.~C.,   Johannes R.~S.,  1988, in Proceedings of the annual symposium on computer application in medical care. p.~261

\bibitem[\protect\citeauthoryear{{Speagle}}{{Speagle}}{2020}]{2020MNRAS.493.3132S}
{Speagle} J.~S.,  2020, \mn@doi [\mnras] {10.1093/mnras/staa278}, \href {https://ui.adsabs.harvard.edu/abs/2020MNRAS.493.3132S} {493, 3132}

\bibitem[\protect\citeauthoryear{Spurio~Mancini, Piras, Alsing, Joachimi  \& Hobson}{Spurio~Mancini et~al.}{2022}]{spurio2022cosmopower}
Spurio~Mancini A.,  Piras D.,  Alsing J.,  Joachimi B.,   Hobson M.~P.,  2022, Monthly Notices of the Royal Astronomical Society, 511, 1771

\bibitem[\protect\citeauthoryear{Spurio Mancini, Docherty, Price  \& McEwen}{Spurio Mancini et~al.}{2023}]{Spurio_Mancini_2023}
Spurio Mancini A.,  Docherty M.~M.,  Price M.~A.,   McEwen J.~D.,  2023, \mn@doi [RAS Techniques and Instruments] {10.1093/rasti/rzad051}, 2, 710–722

\bibitem[\protect\citeauthoryear{Srinivasan, Crisostomi, Trotta, Barausse  \& Breschi}{Srinivasan et~al.}{2024}]{srinivasan2024floz}
Srinivasan R.,  Crisostomi M.,  Trotta R.,  Barausse E.,   Breschi M.,  2024, floZ: Evidence estimation from posterior samples with normalizing flows (\mn@eprint {arXiv} {2404.12294})

\bibitem[\protect\citeauthoryear{Torrado \& Lewis}{Torrado \& Lewis}{2019}]{torrado2019}
Torrado J.,  Lewis A.,  2019, Astrophysics Source Code Library, pp ascl--1910

\bibitem[\protect\citeauthoryear{Trotta}{Trotta}{2008}]{trotta2008bayes}
Trotta R.,  2008, Contemporary Physics, 49, 71

\bibitem[\protect\citeauthoryear{Williams}{Williams}{1959}]{williams:1959}
Williams E.~J.,  1959, Regression analysis.
Wiley, New York

\bibitem[\protect\citeauthoryear{Williams, Veitch  \& Messenger}{Williams et~al.}{2021}]{Williams_2021}
Williams M.~J.,  Veitch J.,   Messenger C.,  2021, \mn@doi [Physical Review D] {10.1103/physrevd.103.103006}, 103

\bibitem[\protect\citeauthoryear{van Haasteren}{van Haasteren}{2014}]{vanhaasteren:2014}
van Haasteren R.,  2014, in , Gravitational Wave Detection and Data Analysis for Pulsar Timing Arrays.
Springer, pp 99--120

\makeatother
\end{thebibliography}

\end{document}